\documentclass[%
 amsmath,amssymb,
 aps,
prf,
]{revtex4-2}

\usepackage{graphicx}
\usepackage{dcolumn}
\usepackage{bm}
\usepackage{color,graphicx}
\usepackage[dvipsnames]{xcolor}
\usepackage{ulem}
\usepackage{epstopdf, epsfig}
\usepackage{subcaption}
\usepackage{dashrule}
\usepackage{multirow}
\usepackage{natbib,url,hyperref}
\usepackage{amsmath}
\usepackage{lipsum}
\usepackage{tabularx}
\usepackage{rotating}
\usepackage{hyperref}
\usepackage{color}
\usepackage{xcolor}

\definecolor{grey}{rgb}{0.7,0.7,0.7}

\definecolor{cNeutralGray}{RGB}{99,101,105}
\definecolor{cGreenCurrent}{rgb}{0.4660,0.6740,0.1880}
\definecolor{cEverGreen}{rgb}{0,0.5,0}
\definecolor{cCyanCurrent}{rgb}{0.3010, 0.7450, 0.9330}
\definecolor{cBlueSecondary}{RGB}{0, 75, 135}
\definecolor{cViolet}{RGB}{204, 0, 204}
\definecolor{cYellow}{RGB}{255, 255, 51}
\definecolor{cOrangeUtility}{RGB}{242, 160, 0}
\definecolor{cRedUtility}{RGB}{183, 49, 44}

\definecolor{cBlueBrand}{RGB}{47, 126, 178}
\definecolor{cVioletCurrent}{rgb}{0.4940, 0.1840, 0.5560}
\definecolor{cBlack}{rgb}{0, 0, 0}
\definecolor{cYellowCurrent}{rgb}{0.9290, 0.6940, 0.1250}

\begin{document}


\title {Reconsiderations about inner layer of wall-bounded flows}

\author{Hassan Nagib}%
\email{nagib@iit.edu}
\affiliation{%
ILLINOIS TECH (IIT), Chicago, IL 60616, USA
}%

\date{\today}
\vspace{30 mm}
\begin{abstract}
Following recent evidence that even ZPG boundary layers do not exhibit a purely logarithmic extended overlap region, reconsideration of recently advanced logarithmic plus linear extended overlap region in wall-bounded flows leads to a revision of the model for the extended overlap region. The significant difference between the two representations is a separation between the inner layer and the extended overlap layer in the coefficient of the logarithmic term into $\kappa_{in}$ and $\kappa_o$, respectively. From a wide range of data examined in wall-bounded flows, the value of $\kappa_{in}$ is universal and equal to $1/2.6$ or in the range $0.38< \kappa_{in} < 0.39$. The value of $\kappa_o$ depends on the pressure gradient imposed by the flow geometry. In regard to the trends of the streamwise normal stress, recent publications concluded that the defect-power model developed from bounded dissipation is in more agreement with experimental data from ZPG boundary layers and pipe flows, as well as DNS data for channel and pipe flows, than the logarithmic model developed with inviscid analysis based on wall-scaled eddies.  For some recent investigations and the entire previous literature on this popular topic, the assessment is made in the overlap region between inner and outer flows, which has limited viscous effects and is essentially inviscid; i.e., $y^+_{in} \gtrapprox 400$ and $Y_{out} \approx 0.45$. This appears to be counterintuitive and deserves further attention. Here, both models are reevaluated using the same data sets from recent investigations in a region closer to the wall but outside the region with viscous stresses exceeding $20\%$ of the total stress; i.e., dominated by viscous effects. It is perplexing that both an inviscid and a viscous model agree equally well with experiments and DNS data in this region closer to the wall.  The parameters extracted in this region for the two models are different from those accepted in the literature for the overlap region farther from the wall. The trends of the two coefficients for each of the models are found to be consistent for both experimental measurements and DNS results in channel, pipe and ZPG boundary layer flows over the range $200 < Re_\tau < 100,000$. This work draws attention to substantial limitations for achieving sufficiently high Reynolds numbers to match the requirements of assumptions in some models of wall-bounded flows.
\end{abstract}

%
\maketitle
\section{Introduction}
Reconsiderations and extensions for two important aspects of wall-bounded turbulent flows and their models of overlap and wall layers presented here follow directly from a recent invited talk at the 2024 meeting of the Division of Fluid Dynamics of the American Physical Society by the current author \cite{nagaps24}.  The first part of that presentation highlighted recent findings on the logarithmic plus linear overlap region of the mean velocity by Monkewitz and Nagib (2023) \cite{mon23} and Baxerras et al. (2024) \cite{bax24}. The second part of the presentation focused on the evaluation of two models for normal stress relations by experimental data over a wide Reynolds number range and is reported in Nagib and Marusic (2024) \cite{nagmar24}. 

In the work of Monkewitz and Nagib (2023) \cite{mon23}, the long-accepted logarithmic overlap region was reevaluated with matched asymptotic analysis incorporating an additional term in the inner expansion. This revealed an extended overlap region with an additional linear term in distance to the wall and a coefficient labeled $S_o$. They examined with indicator functions of mean velocity profiles DNS and experimental data from flows of channel, pipe, and zero-pressure gradient boundary layer.  Their work was subsequently extended by Baxerras et al. (2024) \cite{bax24} to favorable and adverse pressure gradient data from two wind tunnels. The results confirm the non-universality of the von K\'arm\'an constant, reported earlier by Nagib and Chauhan (2008) \cite{variations}, and the coefficient of the additional linear term, $S_0$. Motivated by recent evidence from zero pressure gradient (ZPG) boundary layers, the first part of this paper revisits the recently proposed logarithmic plus linear formulation and presents a revised model for the extended overlap region.

Interest in characterizing the normal stress of turbulent fluctuations and its relationship to large-scale eddies has increased significantly over the last five to ten years. Recently, Nagib et al. (2024) \cite{nag24} utilized indicator functions of streamwise normal stress profiles to reveal the validity ranges, in wall distance and Reynolds number, for each of two proposed models in Direct Numerical Simulations (DNS) of channel and pipe flows.  The indicator function approach was also tried for the experimental data by Monkewitz (2023) \cite{M23}. Nagib and Marusic (2024) \cite{nagmar24} proposed a method more suited to experimental data, as establishing accurate indicator functions is a challenge. The new method is outlined and used with the two leading models that propose either a logarithmic or a defect-power trend, for the normal stresses of turbulence in a ``fitting region" of wall-bounded flows.

Two models have been proposed to represent the trends for normal stresses of turbulence in wall-bounded flows, and they are the focus of recent research and numerous publications. The current work is also motivated by recent findings regarding an extended overlap region of the mean velocity profile for wall-bounded turbulent flows by Monkewitz and Nagib (2023) \cite{mon23}, revealing that not all such flows exhibit a pure logarithmic profile and a linear term of the same order should be considered. One potential implication of the additional linear term in the mean flow is that it would require eddies that do not strictly scale with their distance from the wall in this overlap region, as in the wall-scaled eddies model inspired by ``Townsend attached eddies concept" and outlined by Marusic and Monte (2019) \cite{mar19}. Here, this model is referred to as the ``wall scaled eddies model" or simply the ``logarithmic trend" model. For the streamwise normal stress, the trend based on this model is given by:

\begin{equation}\label{eq:001}
     \left\langle u^+u^+\right\rangle (Y) = B_1 - A_1\ln(Y).
\end{equation}

\noindent The outer-scaled distance from the wall $Y$ is defined as $y/\delta$ in the ZPG data and as $y/R$ for the pipe flow, where $\delta$ is the boundary layer thickness and $R$ the pipe radius. The $\delta$ used here is the same as that used by Samie et al. (2018) \cite{sam18} and is equivalent to the one defined by Baxerras et al. (2024) \cite{bax24}.

Recent publications by Chen and Sreenivasan (2022, 2023) \cite{che22,che23} on bounded dissipation introduce an alternative model, which is referred to here as the ``defect-power trend" model. This model is represented by the following relation for the streamwise normal stress:

\begin{equation}\label{eq:002}
    \left\langle u^+u^+\right\rangle (Y) = \alpha_1 - \beta_1~(Y)^{0.25}.
\end{equation}

Both models contain two parameters, and in addition the defect-power trend and its exponent are based on the bounding of dissipation near the wall at infinite Reynolds number. The subscript ``$1$" represents the streamwise component of the normal stresses of turbulence. Analyzing velocity spectra obtained in pipe flow using direct numerical simulations, Pirozzoli (2024) \cite{piroz24} suggested that the dissipation rate of the streamwise velocity reaches a limiting value of $0.28$ for high $Re_\tau$. Hence, it is worth testing parametrically for the power of the exponent in equation \ref{eq:002}.

Agreement with the logarithmic trend is often evaluated in the literature, particularly in comparisons with experimental data, by fitting a straight line over a segment of normal stress data on a semi-log plot in the overlap region of wall-bounded flows; examples are found in Marusic et al. (2013) \cite{mar13}, and more recently in Hwang et al. (2022) \cite{hwa22} and Diwan and Morrison (2021) \cite{Diw21}.  This approach makes it challenging to assess the accuracy and ranges of validity of the model, particularly due to the often sparse logarithmic spacing of results in the fitting and outer regions, as well as the limited accuracy of experimental data. For the trend of $0.25$ power, agreement with the data is typically tested by iteratively adjusting the two parameters $\alpha_1$ and $\beta_1$, seeking minimum deviations over the widest possible range of distances from the wall in a fitting region between the walls and the outer flows. Equations \ref{eq:001} and \ref{eq:002} are independent of the Reynolds number and $Re_\tau$ enters into them through the inner limit of the fitting procedure in $y^+$. This is the main reason $Y = y^+ / Re_\tau$ is used here for the distance to the wall almost exclusively.

In the case of DNS data, indicator functions analogous to those utilized with mean velocity profiles in Monkewitz and Nagib (2023) \cite{mon23} were identified as having the best potential for normal stress analysis and were favored by Nagib et al. (2024) \cite{nag24}. However, just like with mean velocity, these indicator functions require taking derivatives of the discrete normal-stress profiles.  This approach is often unsuitable for most experimental data in the literature due to limited spatial resolution and experimental accuracy, which hamper the ability to obtain accurate derivatives of the profiles.  The work of Nagib and Marusic (2024) \cite{nagmar24} aimed to develop a method to evaluate the proposed relations using two of the most well documented and reliable experimental results in the boundary layers of zero pressure gradient and pipe flows, as provided by Samie et al. (2018) \cite{sam18} and Hultmark et al. (2012) \cite{hul12}, respectively.

The simple and robust method can be effectively used to evaluate each model over a wide range of Reynolds numbers ($Re_\tau$) by applying it to experimental data sets from any wall bounded flow. For the two data sets evaluated by Nagib and Marusic (2024) \cite{nagmar24}, significant regions of validity of either relation are only found when $Re_\tau \gtrapprox 10,000$, with the lower limit $y^+_{in} \sim (Re_\tau)^{0.5}$ and the provision that $y^+_{in} \gtrapprox 400$ for the ZPG and pipe flow data. The upper limit is found to be a fixed fraction of the thickness of the boundary layer or the radius of the pipe, and is independent of $Re_\tau$. However, the outer limit of validity for the defect-power trend is approximately twice that of the logarithmic model and is identified at $Y \approx 0.4$ for ZPG and $Y \approx 0.5$ for the pipe flow data. The initial evaluations of the defect-power trend model were made by Nagib and Marusic (2024) \cite{nagmar24} with the exponent of $0.25$ based on the bounded dissipation results.  They also tested other exponents in the range from $0.2$ to $0.32$. For ZPG boundary layers, data from the Melbourne large wind tunnel by Samie et al. (2018) \cite{sam18} and Marusic et al. (2015) \cite{mar15} were used, with an emphasis on recent results using more advanced hot wire anemometers that provided better spatial resolution. Similarly for the normal stress data from the Princeton Superpipe facility, the more recent data from NSTAP hot wire sensors with a shorter sensor length from Hultmark et al. (2012) \cite{hul12} were used.

 The two parameters for both models of equations \ref{eq:001} and \ref{eq:002} are found to be similar for the two flows, and using the same parameters for the full range of $Re_\tau$ in each flow results in a lower limit of wall distance independent of Reynolds number. For both flows and models, the lower limit in wall distance is $y^+_{in} \gtrapprox 400$, which corresponds to $0.004 \lessapprox Y_{in} \lessapprox 0.04$ for the range of Reynolds numbers for both measurements in the ZPG boundary layer and the Superpipe examined. However, the outer limit of validity for the defect-power trend is nearly double that of the logarithmic trend and is identified at $Y_{out} \approx 0.39$ versus $\approx0.22$ for ZPG and $Y_{out} \approx 0.5$ versus $\approx0.27$ for Superpipe data. Both models are not valid outside of this fitting region identified between $y^+_{in}$ and $Y_{out}$. The standard deviation from the mean values of the two parameters for each of the proposed relations was found to be quite small for all data analyzed and did not exceed $2.5\%$ of the mean values of the respective parameters. The defect-power trend, which is based on Chen and Sreenivasan (2021) \cite{che21}, has an explicit formulation of the inner region, but that is not the subject of this paper. 

Nagib and Marusic (2024) \cite{nagmar24} conclude that while both models may be used to represent the general trends of the streamwise normal-stress data in the overlap region, the defect-power relation conforms to the experimental data more closely throughout its wider range of validity. 
They also find that a somewhat larger exponent for the defect-power law of streamwise normal stress equal to $0.28$, instead of $0.25$ obtained by the bounded dissipation assumption of Chen and Sreenivasan (2022, 2023) \cite{che22} and \cite{che23}, is found to extend its validity range. Also, recognizing that the outer part of boundary layers is intermittent between laminar and turbulent conditions,  a correction to the normal stress is applied by dividing it by the intermittency factor as representative of the fraction of time the flow is turbulent.  The resulting validity of the defect-power trend is extended from $Y_{out} \approx 0.4$ to around half the thickness of the boundary layer. 

In the following sections and in a similar way to the presentation of Nagib (2024) \cite{nagaps24}, newer insights developed regarding mean flow are presented first. They clarify the ranges of validity of the pure logarithmic (pure-log) region and the extended overlap region with its logarithmic plus linear (log+lin) trend, and their correlation with the constant turbulence stress region. Next, the log+Lin extended overlap model of Monkewitz and Nagib (2023) \cite{mon23} is revisited with a focus on a better characterization of the inner region for the model. Finally, after reevaluating the extent of the viscous dominated part of the inner or wall region based on recent DNS data, the two normal stress models are evaluated in a different region of three wall-bounded flows closer to the inner peak of the normal stresses rather than in the overlap region, as carried out by Nagib and Marusic (2024) \cite{nagmar24}. The three flows included DNS data in channel and pipe flows and experimental data in zero pressure gradient boundary layers and pipe flows.

\begin{figure}
      \centering
      \includegraphics[width=0.7\textwidth]{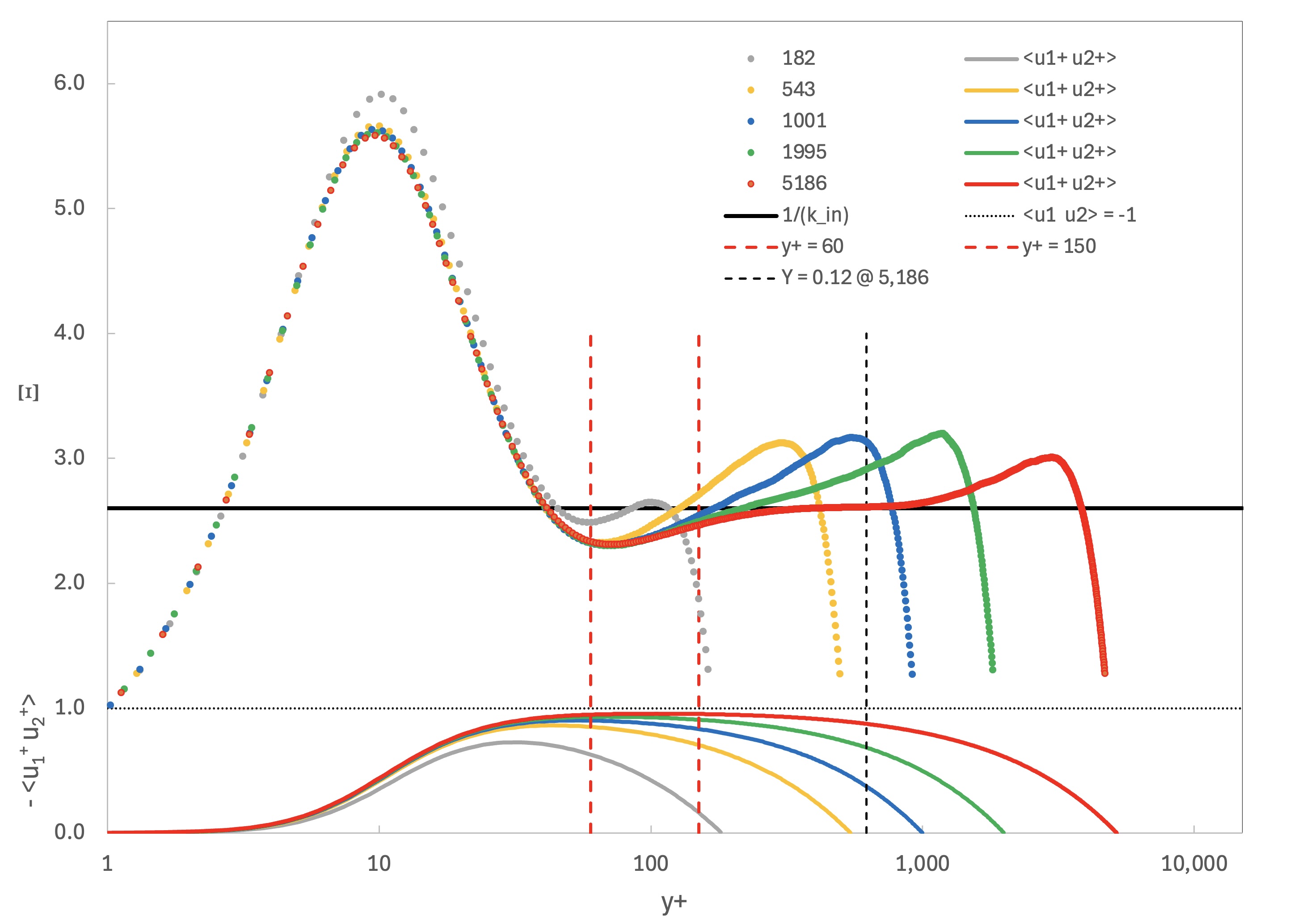}
       \caption{Indicator function of mean velocity, $\Xi$, and turbulent stress, $<u_1^+ u_2^+>$, versus inner-scaled wall distance, $y^+$, for different $Re_\tau$'s of channel flow from DNS results by Lee and Moser (2015) \cite{lee15}.}
       \label{fig:fig1}
\end{figure}

\begin{figure}
      \centering
      \includegraphics[width=0.7\textwidth]{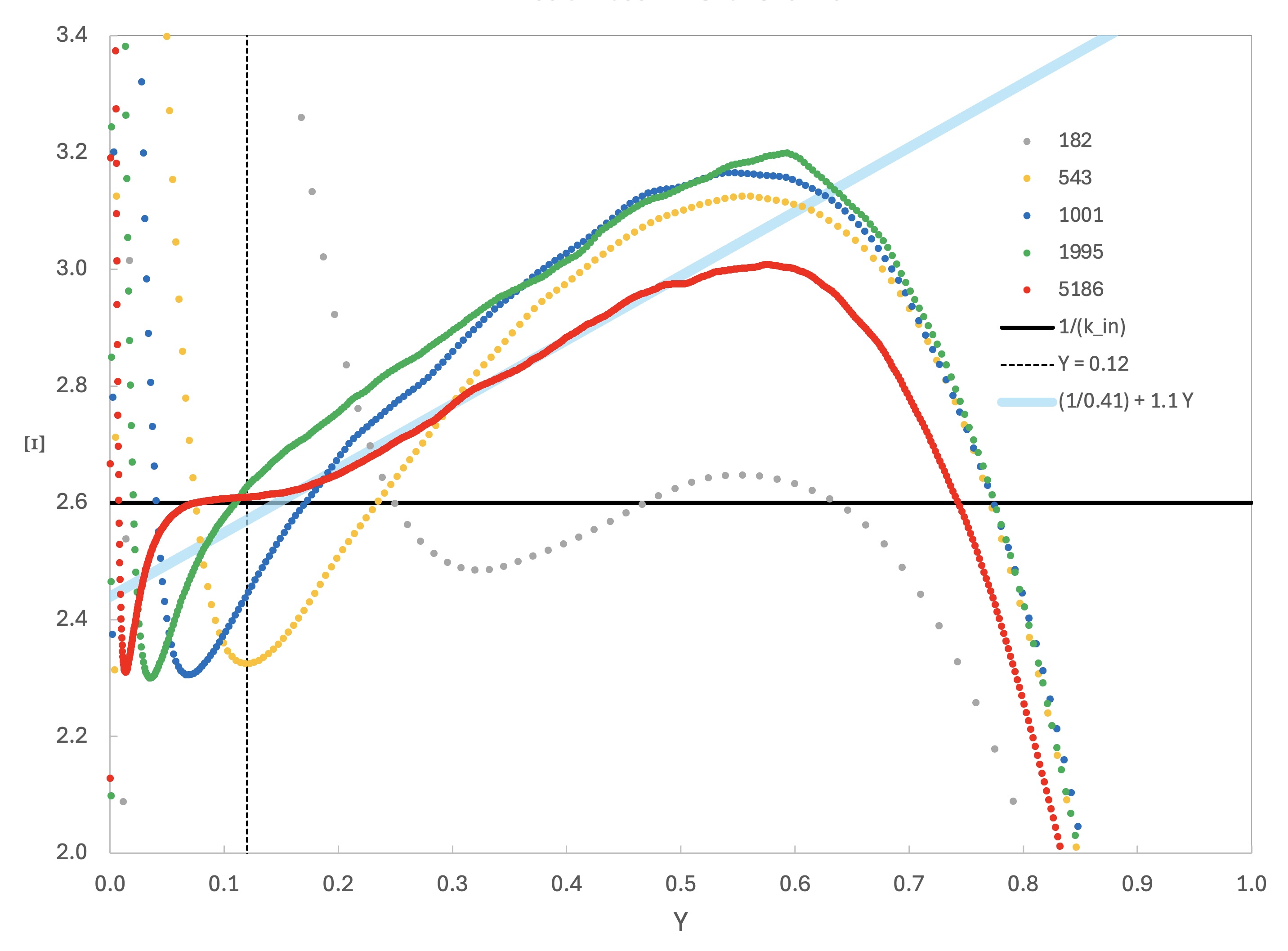}
        \caption{Indicator function of mean velocity, $\Xi$, versus outer-scaled wall distance, $Y$, for different $Re_\tau$'s of channel flow from DNS results by Lee and Moser (2015) \cite{lee15}.}
        \label{fig:fig2}
\end{figure}

\begin{figure}
      \centering
      \includegraphics[width=0.7\textwidth]{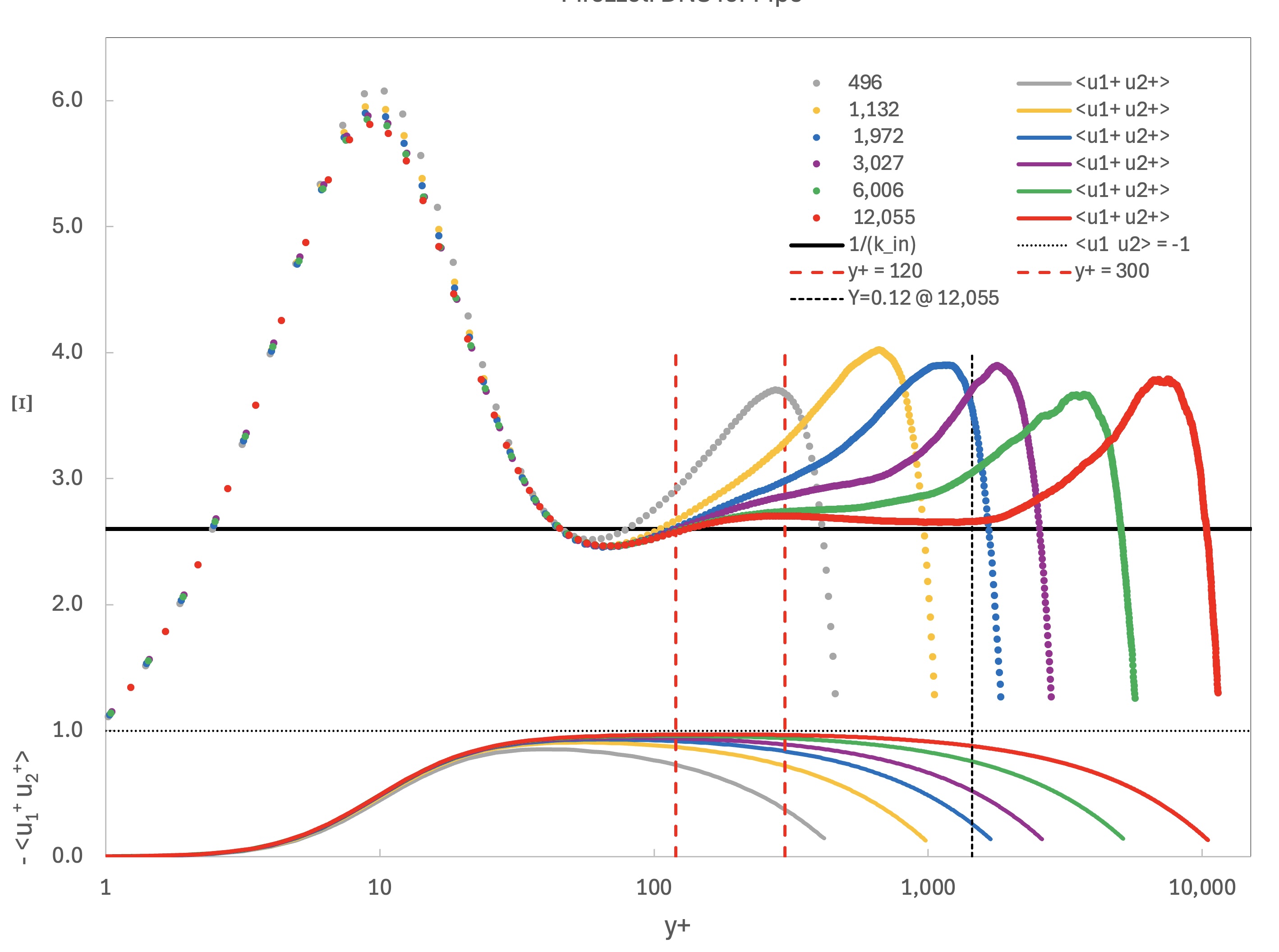}
        \caption{Indicator function of mean velocity, $\Xi$, and turbulent stress, $<u_1^+ u_2^+>$, versus inner-scaled wall distance, $y^+$, for different $Re_\tau$'s of pipe flow from DNS results by Pirozzoli (2024) \cite{piroz24}.}
       \label{fig:fig3}
\end{figure}

\begin{figure}
      \centering
      \includegraphics[width=0.75\textwidth]{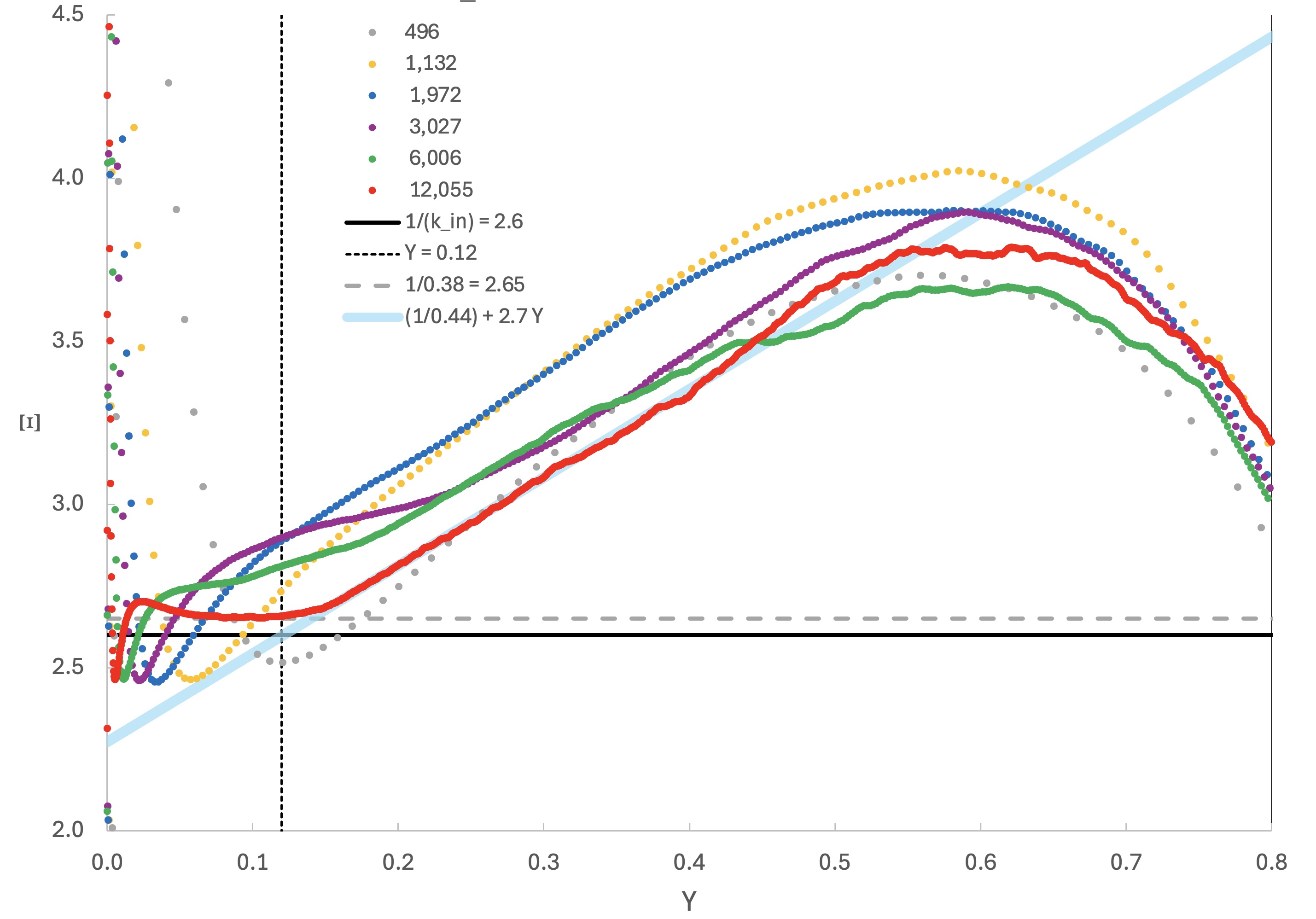}
        \caption{Indicator function of mean velocity, $\Xi$, versus outer-scaled wall distance, $Y$, for different $Re_\tau$'s of pipe flow from DNS results by Pirozzoli (2024) \cite{piroz24}.}
        \label{fig:fig4}
\end{figure}

\section{Indicator functions of mean velocity and the uniform stress layer}
The pure-log relation for an overlap region of the mean velocity profile follows equation (\ref{eq:003}), which contains $\kappa$, the von K\'arm\'an constant. 
\begin{equation}\label{eq:003}
    \overline{U}^+_{1\rm OL}(y^+\gg 1~\&~Y\ll 1) = \frac{1}{\kappa} \ln y^+ + B.
\end{equation}
In this equation, $\overline{U}_1$ is the streamwise mean velocity, the ``${\rm +}$" superscript indicates that viscous scaling is being used and the ``${\rm OL}$" subscript indicates validity in the overlap region. The independent variable $y^+$ is the inner-scaled wall normal coordinate, $y^+ = y u_\tau /\nu$, where $u_\tau$ is the friction velocity and $\nu$ is the fluid kinematic viscosity. The variable $Y$ is the normal wall coordinate made dimensionless by the ``outer scale of the flow" $\ell$. 
Recently, Monkewitz and Nagib (2023) \cite{mon23} shed additional light on this topic, challenging the knowledge accumulated over the last century. The main point of their work is that a more detailed matched asymptotic expansion (MAE) of wall-bounded flows produces an overlap region containing a combination of logarithmic and linear terms of the distance to the wall as shown in equation (\ref{eq:004}). They arrived at this result by considering in the inner asymptotic expansion a term proportional to the wall normal coordinate, $\mathcal{O}(Re_{\tau}^{-1})$. Here, this overlap region is referenced to as ``log+lin", with $Re_{\tau} = u_\tau \ell / \nu$, which is the friction Reynolds number, and $\ell$ a characteristic outer length. This $\ell$ can be thought of as the radius $R$ in pipes, the semi height in channels, $h$, or a length scale related to the thickness 99\% of the boundary layer, $\delta_{99}$, in boundary layers. 
\begin{equation}\label{eq:004}
{U_1}^+_{\rm OL}(Y\ll 1) = \kappa^{-1}\ln Y + \kappa^{-1}\ln Re_\tau +S_o  Y + B_o + {\rm  H.O.T.}.
\end{equation}

The commonly used indicator function for the mean velocity profile, $\Xi$, can be obtained by:
\begin{equation}\label{eq:005}
 \Xi = y^+\frac{{\rm d}{U}_1^+}{{\rm d}y^+} = Y\frac{{\rm d}{U}_1^+}{{\rm d}Y}.   
\end{equation}
From equations (\ref{eq:004}) and (\ref{eq:005}), one can obtain the equation for $\kappa$ and $S_o$:
\begin{equation}\label{eq:006}
\Xi_{\rm OL} = \kappa^{-1} +S_o  y^+ /Re_\tau = \kappa^{-1} +S_o  Y .
\end{equation}

The indicator function, $\Xi$, calculated from the DNS data in channel flow by Lee and Moser (2015) \cite{lee15} under different $Re_\tau$ conditions, is displayed in the upper part of figure \ref{fig:fig1} as a function of the inner scaled wall distance $y^+$.  In the lower part of the figure, the turbulence shear stress $<u_1^+ u_2^+>$ is shown for the various cases of $Re_\tau$ using the same colors for each Reynolds number. Here, we represent the velocity component of turbulence in the streamwise and wall normal directions with $u_1$ and $u_2$, respectively, and the brackets <..> indicate the time average.  For the highest $Re_\tau$, the region of nearly constant turbulence stress is marked with two vertical dashed red lines. It is interesting to note that the only pure-log region in this entire high-quality data set is found for the highest $Re\_tau$ and extends starting from beyond the constant stress region up to the vertical dashed black line, or from near $y^+ = 150$ to approximately $Y = 0.12$.

An enlarged presentation of the same indicator function data plotted versus the outer scaled wall distance $Y$ is shown in figure \ref{fig:fig2}, and the best fit of equation \ref{eq:006} for the highest $Re_\tau$ case is shown with a light blue line. The values of $\kappa$ and $S_o$  for this case are $0.41$ and $1.1$, respectively.  The dashed black line of figure \ref{fig:fig1} is shown again in this figure.

In figures \ref{fig:fig3} and \ref{fig:fig4}, the DNS data of Pirozzoli (2024) \cite{piroz24} for pipe flow are shown in the same way as in figures \ref{fig:fig1} and \ref{fig:fig2}. Although a pure-log region is not sufficiently defined here for either of the highest $Re_\tau$ cases, the constant shear stress region for each case does not represent a constant $\Xi$ region. Both figures \ref{fig:fig1}and \ref{fig:fig3} strongly refute the often cited derivation of the pure-log law based on a constant stress region. For the highest case of $Re_\tau$ in the flow of the pipe, the best fit for the extended overlap region, shown with the light blue line, and the values of $\kappa$ and $S_o$ are $0.44$ and $2.7$, respectively.  Compared to the results for the channel flow presented above, it is clear that the overlap parameters are not universal in wall-bounded flows, as originally proposed by Nagib and Chauhan (2008) \cite{variations} and confirmed more recently by Monkewitz and Nagib \cite{mon23} and Baxerras et al. \cite{bax24}.

\begin{figure}
      \centering
      \includegraphics[width=0.65\textwidth]{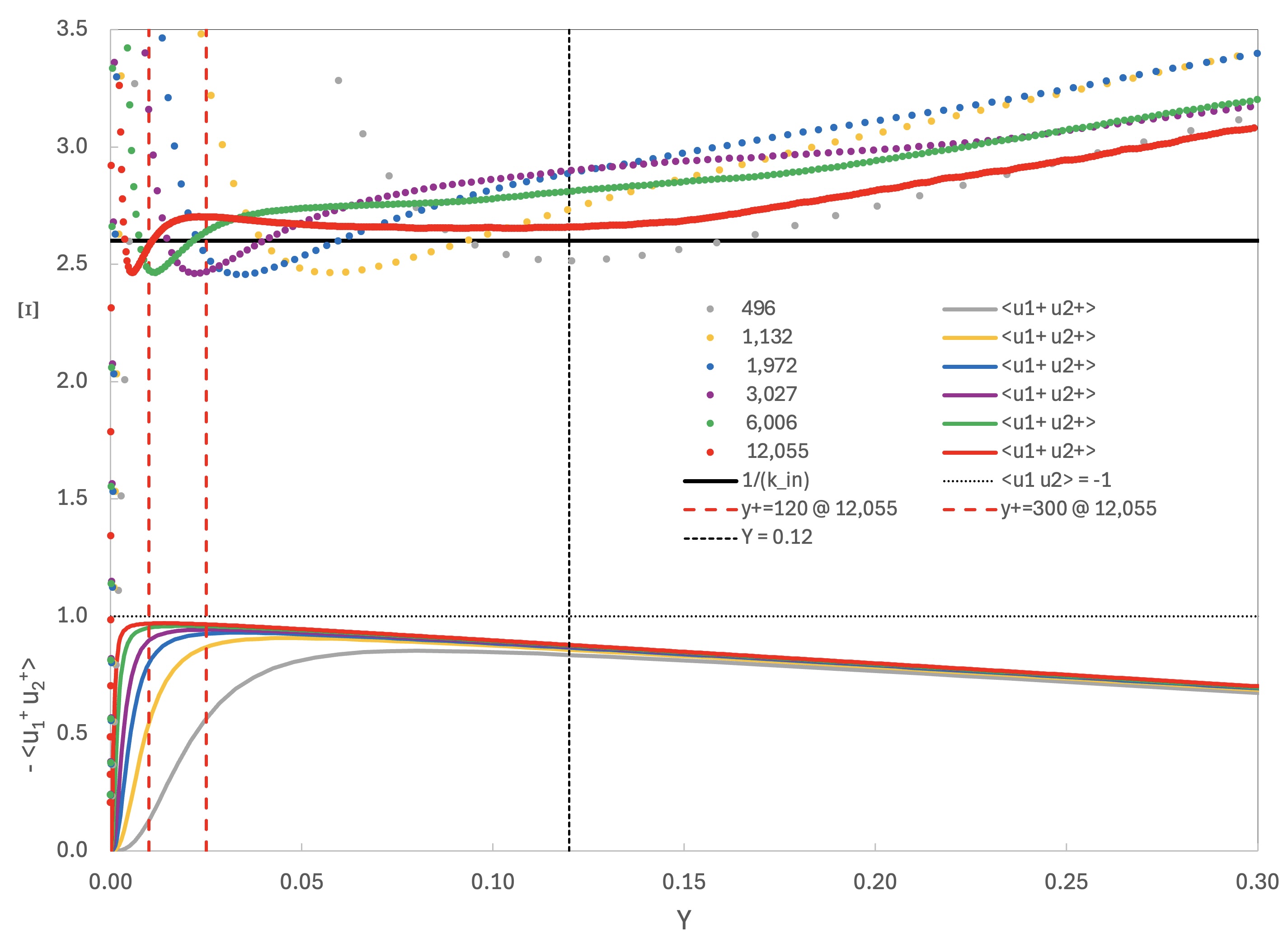}
        \caption{Indicator function of mean velocity, $\Xi$, and turbulent stress, $<u_1^+ u_2^+>$, versus outer-scaled wall distance, $Y$, for different $Re_\tau$'s of pipe flow from DNS results by Pirozzoli (2024) \cite{piroz24}.}
        \label{fig:fig5}
\end{figure}

Again, an enlarged presentation of the same data shown in figure \ref{fig:fig3} plotted versus the outer scaled wall distance $Y$ is shown in figure \ref{fig:fig5}. The lack of correspondence between the best pure-log region for the highest $Re_\tau = 12,055$, with a constant $\Xi$ and a nearly constant segment of $<u_1^+ u_2^+>$ refutes the derivation of a logarithmic overlap region in wall-bounded flow based on a constant stress layer.

\begin{figure}
      \centering
      \includegraphics[width=0.7\textwidth]{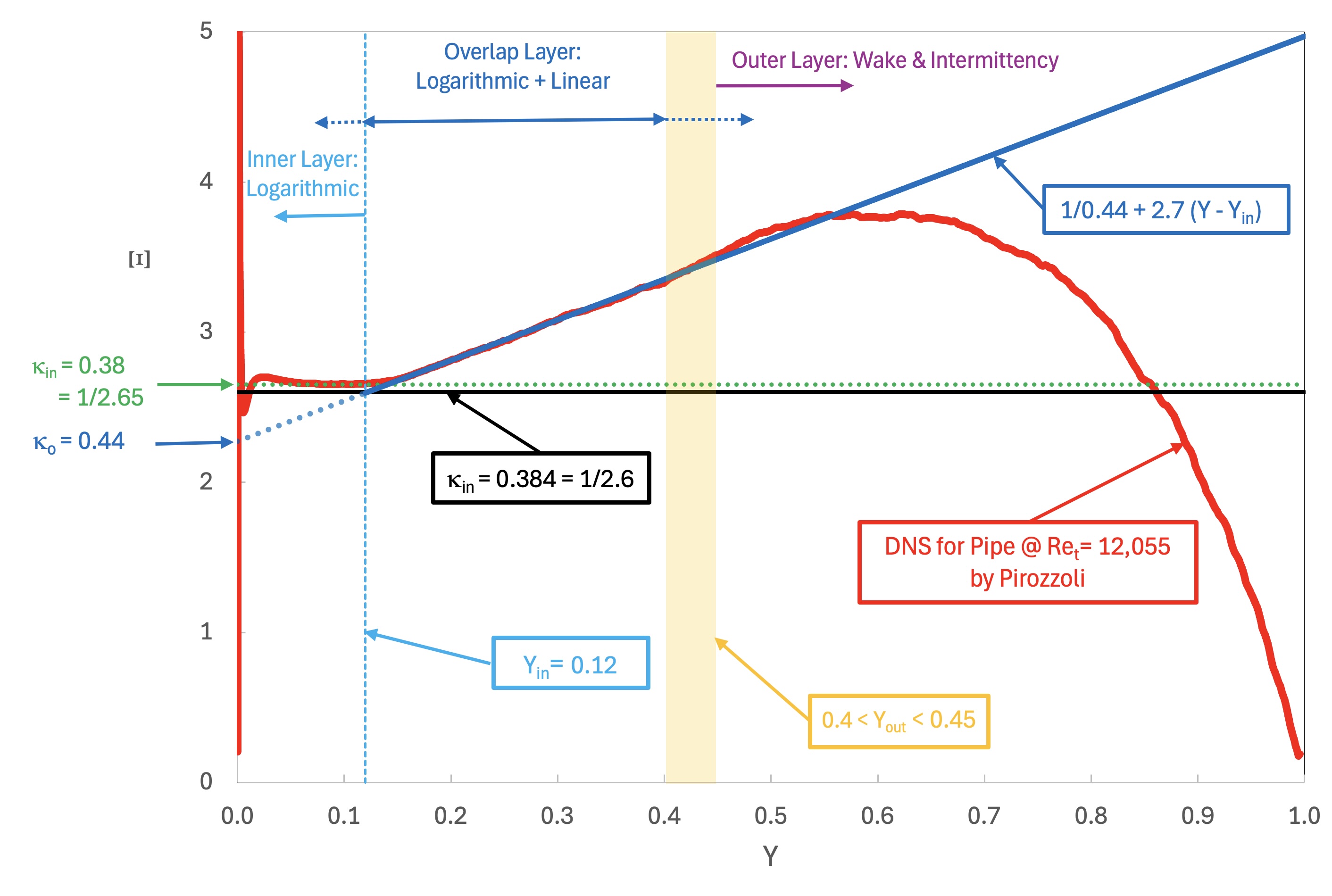}
        \caption{Depiction in indicator function , $\Xi$, versus outer-scaled wall distance, $Y$, of a logarithmic plus linear overlap model between pure logarithmic inner layer and an outer ``wake'' layer.}
        \label{fig:fig6}
\end{figure}

\begin{figure}
      \centering
      \includegraphics[width=0.6\textwidth]{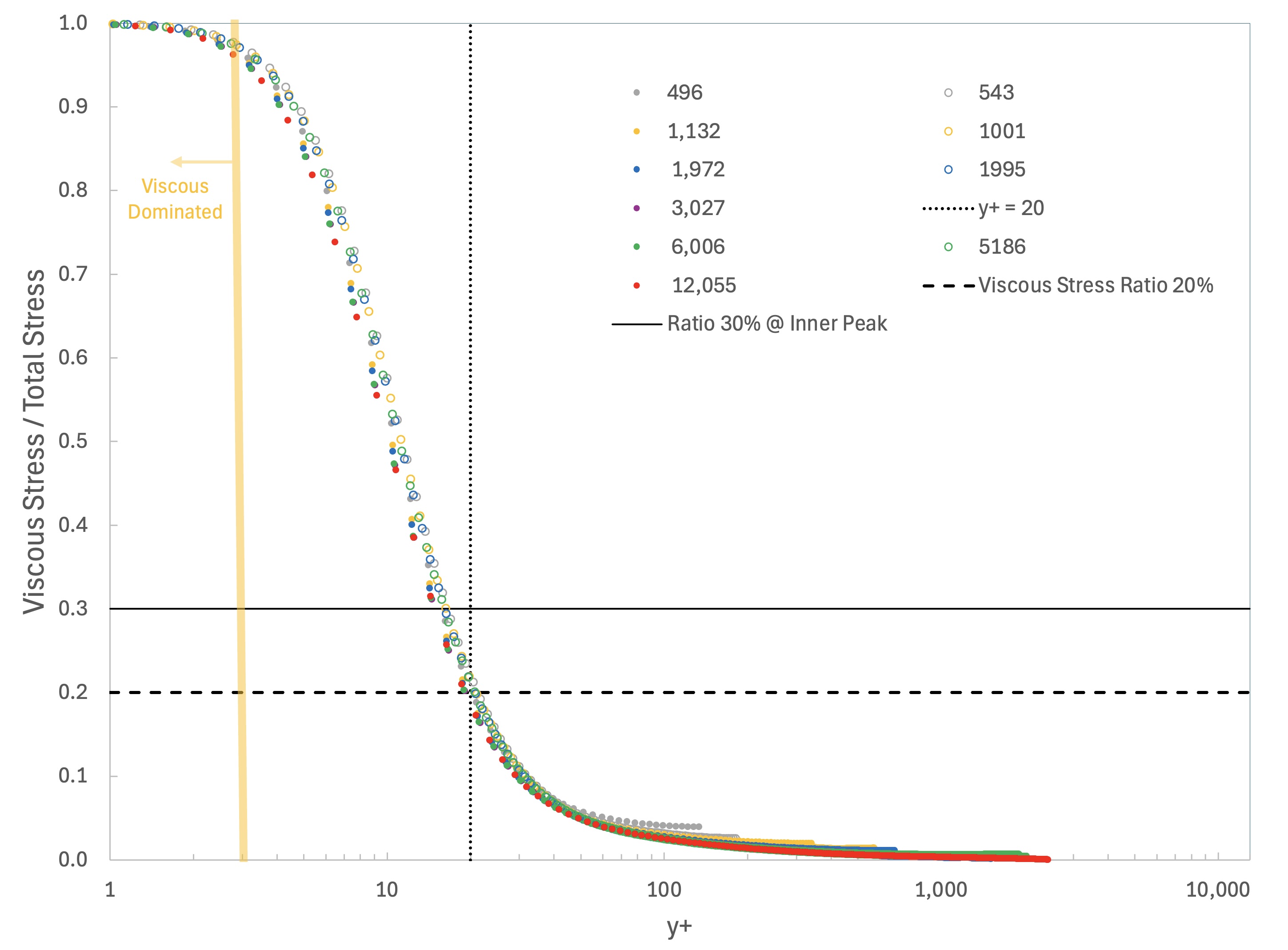}
        \caption{Ratio of viscous to total stress versus inner-scaled wall distance, $y^+$, for different $Re_\tau$'s from DNS results for channel flow by Lee and Moser (2015) \cite{lee15} and pipe flow by Pirozzoli (2024) \cite{piroz24}, shown with open and closed symbols, respectively.}
        \label{fig:fig7}
\end{figure}

\section{Revised logarithmic plus linear overlap model of mean velocity}
In the work of Monkewitz and Nagib (2023) \cite{mon23} challenges were encountered with the few boundary layer cases which were all for zero pressure gradient (ZPG) conditions.  This is revealed in their figure 7, where the pure-log region extended up to only $Y=0.11$. Beyond that region, a linear term is clearly present and was assumed to be part of the wake that represents the outer flow.  In a large part of the literature, experiments in boundary layers, pipes, and channels revealed only logarithmic or nearly logarithmic trends up to $0.12 < Y < 0.15$. Baxerras et al. (2024) \cite{bax24} considered a larger collection of high-quality ZPG boundary layer with independent shear stress measurements with accurate oil film interferometry, among many cases of adverse and favorable pressure gradient boundary layers and pipe and channel flows.  Figure 5 of Baxerras et al \cite{bax24} includes several cases at or near ZPG conditions marked with light gray ``x" symbols representing boundary layer overlap regions with non-zero linear term parameter, $S_o$.  This unexplained trend suggests that even ZPG boundary layers that were considered to have a pure-log extended overlap by Monkewitz and Nagib (2023) \cite{mon23}, Nagib and Chauhan (2008) \cite{variations}, and numerous others in the literature may not have a pure-log extended overlap. 

Alternatively, the pure-log region in the experimental evidence from ZPG boundary layers and from the highest $Re_\tau$ cases of channel and pipe flow DNS results shown in figures \ref{fig:fig1} through \ref{fig:fig6} may be considered to represent the ``inner layer" for the extended overlap region of equation \ref{eq:006}. The revised model of such an inner layer with a pure-log mean velocity profile is established at sufficiently high values of $Re_\tau \gtrapprox 10,000$ and coexists with the viscous-dominated wall region exhibiting a linear velocity profile for $y^+ \lessapprox 3$; see figure \ref{fig:fig7}.  In this model, this inner layer is assumed to be universal among all wall-bounded turbulent flows. Additional matched asymptotic work is now under consideration, but is not covered here. Instead, a slightly revised version of equation \ref{eq:006} is proposed to further fit all wall-bounded turbulent flows and in particular boundary layers.  The large outer ``wake" flow in boundary layers and especially under adverse pressure gradient conditions requires this revision as shown in figure \ref{fig:fig6} and by the following equation:
\begin{equation}\label{eq:007}
\Xi_{\rm OL} = \kappa_o^{-1} +S_o  (y^+ - y_{in}^+)/Re_\tau = \kappa^{-1} +S_o  (Y - Y_{in}) .
\end{equation}

In figure \ref{fig:fig6} the highest $Re_\tau$ data from Lee and Moser (2015) \cite{lee15} are used for illustration and a value of $Y_{in} = 0.12$ is established.  Current work with experimental data from boundary layers under various pressure gradients is parametrically examining the best values for $Y_{in}$, which may be in the range from $0.11$ up to $0.15$, and flow type or pressure gradient dependent. Similarly, $Y_{out}$ is expected to have some variations depending on the type of flow and even more with pressure gradient. In the example of figure \ref{fig:fig6}, the wake component is very small for a channel flow. For pipe flow, it is somewhat larger, increasing considerably for boundary layers and especially in the presence of an adverse pressure gradient. A deeper understanding of such a dependence is also important to study.

Although this data set reveals a value for $\kappa_{in} = 0.384$, or $1/\kappa_{in} = 2.65$, it is proposed based on all measurements from various wall-bounded flows at high $Re_\tau$ conditions examined, that the best value to use for the coefficient of the limited range of  pure-log layer is a universal value for all wall-bounded flows and given by $1/\kappa_{in} = 2.6$.  The value of $\kappa_o$ extracted from the log+lin part of the extended overlap region is not universal and depends on the pressure gradient associated with the flow geometry.  This is consistent with the result $0.38 < \kappa_{in} < 0.39$ found for ZPG boundary layers by Monkewitz and Nagib (2023) \cite{mon23}, Samie et al (2018) \cite{sam18}, Nagib and Chauhan (2008) \cite{variations}  Chauhan et al. (2009) \cite{FDR09}, Baxerras et al. (2024) \cite{bax24}, and many other recent publications.  This figure is intended as a representation of the current reconsideration, and proposed model, of the log+lin extended overlap for the mean velocity of wall-bounded flows. 

\section{Evaluation of streamwise normal stress models closer to the wall}
The comparisons of the logarithmic and defect-power models for normal stress profiles in wall-bounded flows by Nagib and Marusic (2024) \cite{nagmar24} and many other publications, including several referenced here, have generally focused on the overlap region. This overlap region typically lies between the lower limit in wall distance of $y^+_{in} \gtrapprox 400$, which corresponds to $0.04 \lessapprox Y_{in} \lessapprox 0.004$ and $Y_{out} \approx 0.45$.  Both models are not valid with typical values of the parameters $A_1$ and $B_1$ of the equation \ref{eq:001} and $\alpha_1$ and $\beta_1$ of the equation \ref{eq:002} outside of this fitting region identified between $y^+_{in}$ and $Y_{out}$. The extensive work based on bounded dissipation includes matched asymptotic analyzes leading to inner and outer layer profiles of normal stresses as reported by Chen and Sreenivasan (2021) \cite{che21}. The defect-power trend is found in the overlap region between the two layers. 

The logarithmic trend is based on an inviscid model based on wall-scaled eddies. To find an alternate region of wall-bounded flows other than the overlap region examined by Nagib and Marusic \cite{nagmar24} where the conditions for the logarithmic model and the defect-power model are satisfied, the ratio of viscous to total stresses may provide a clue.  Although some experimental measurements are available for this purpose, well-resolved and converged DNS, data are a better choice; e.g., see Hoyas et al. (2024) \cite{hoyas24}. 
\begin{figure}
      \centering
      \includegraphics[width=0.7\textwidth]{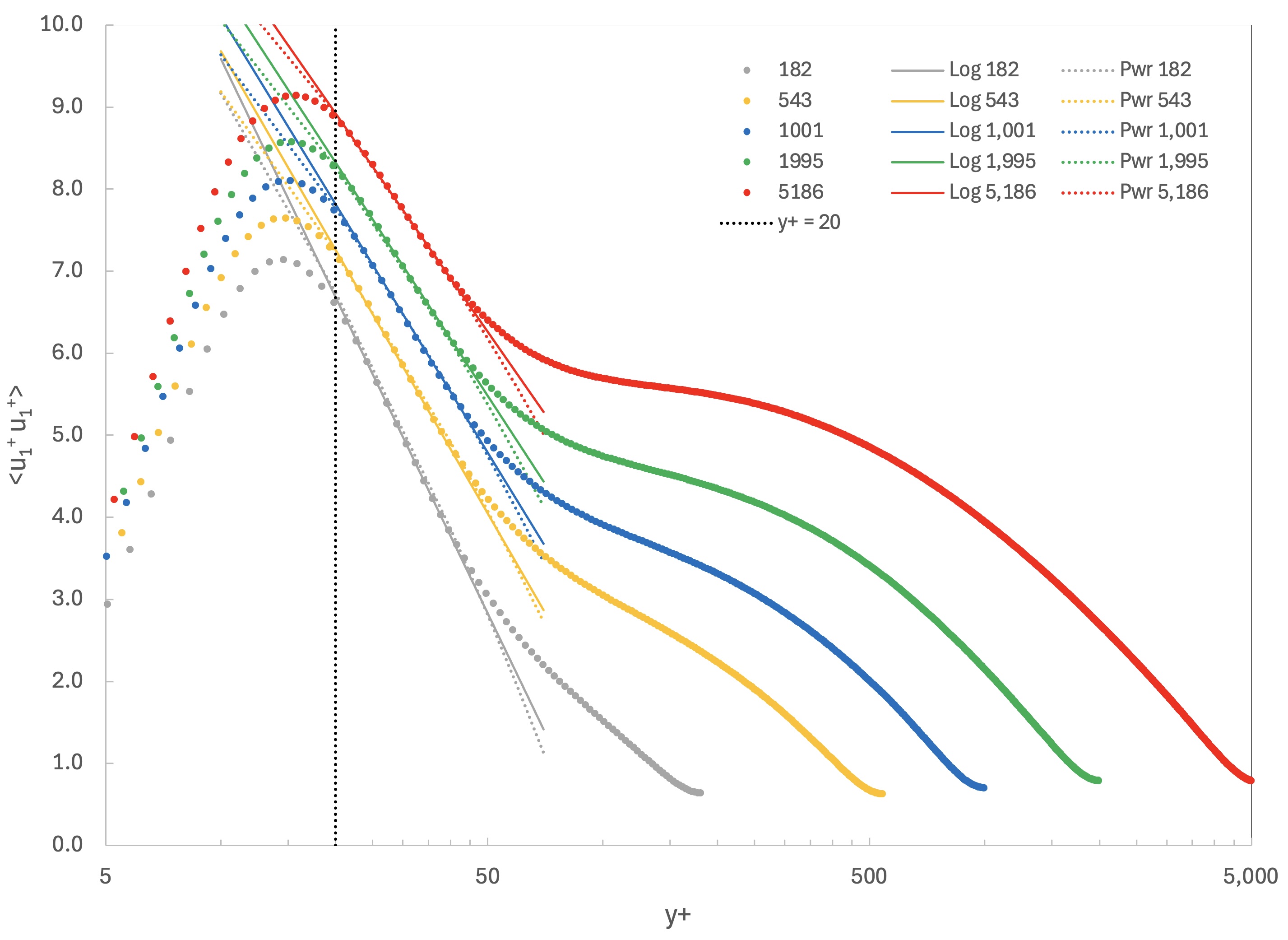}
        \caption{Normal stress, $<u_1^+ u_1^+>$, versus inner-scaled wall distance, $y^+$, for different $Re_\tau$'s of channel flow from DNS results by Lee and Moser (2015) \cite{lee15}, and corresponding best fits of region just outside inner peak by the logarithmic and defect-power models, using solid and dotted lines, respectively.}
        \label{fig:fig8}
\end{figure}

\begin{figure}
      \centering
      \includegraphics[width=0.7\textwidth]{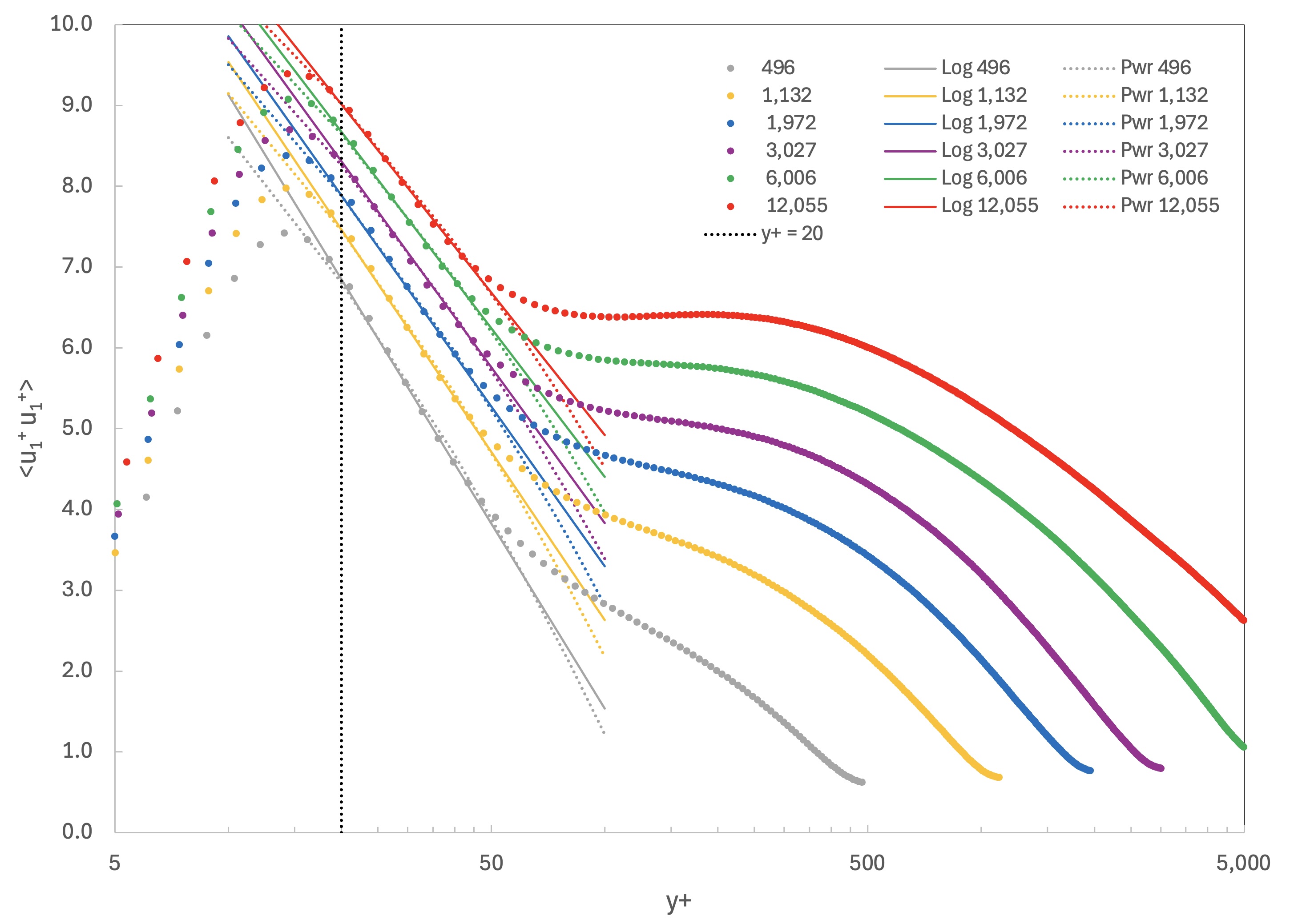}
        \caption{Normal stress, $<u_1^+ u_1^+>$, versus inner-scaled wall distance, $y^+$, for different $Re_\tau$'s of pipe flow from DNS results by Pirozzoli (2024 \cite{piroz24}, and corresponding best fits of region just outside inner peak by the logarithmic and defect-power models, using solid and dotted lines, respectively.}
        \label{fig:fig9}
\end{figure}

\begin{figure}
      \centering
      \includegraphics[width=0.75\textwidth]{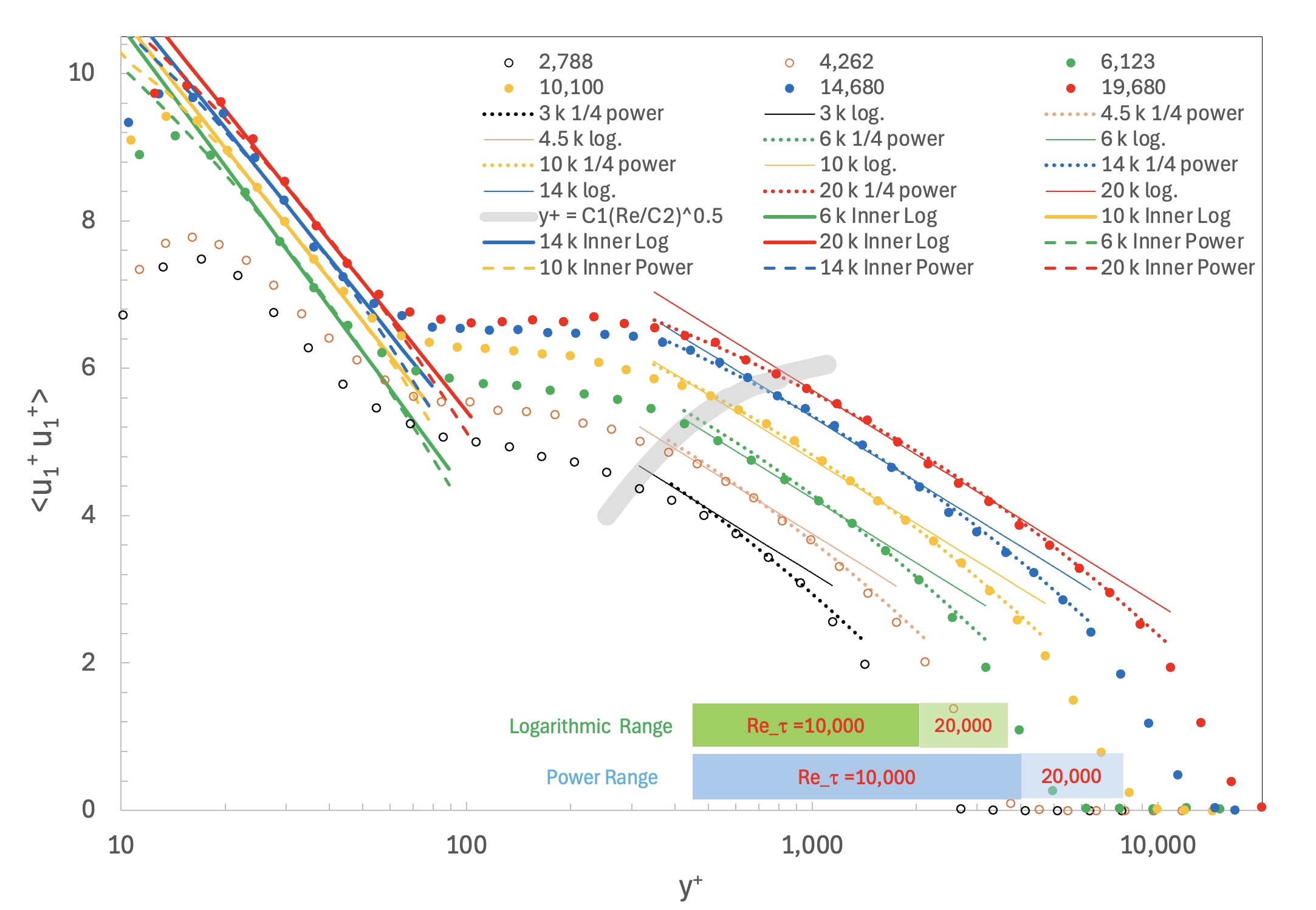}
        \caption{Normal stress, $<u_1^+ u_1^+>$, versus inner-scaled wall distance, $y^+$, for different $Re_\tau$'s of experimental results from ZPG boundary layer by Samie et al. (2018 \cite{sam18}, and corresponding best fits in both the region just outside inner peak and in the overlap region by the logarithmic and defect-power models, using solid and dotted lines, respectively.}
        \label{fig:fig10}
\end{figure}

\begin{figure}
      \centering
      \includegraphics[width=0.75\textwidth]{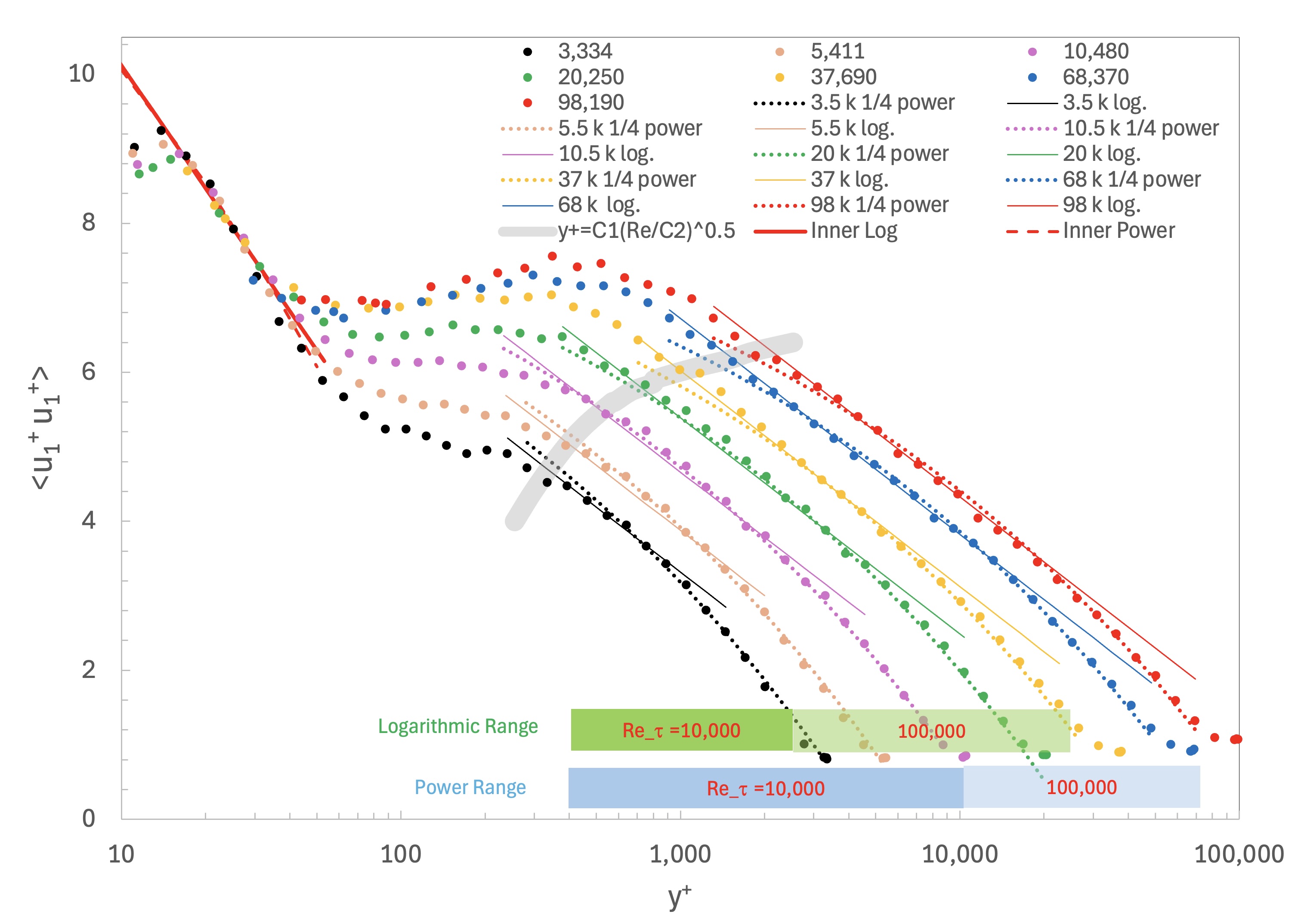}
        \caption{Normal stress, $<u_1^+ u_1^+>$, versus inner-scaled wall distance, $y^+$, for different $Re_\tau$'s of experimental results from pipe flow by Hultmark et al. (2012) \cite{hul12}, and corresponding best fits  in both the region just outside inner peak and in the overlap region by the logarithmic and defect-power models, using solid and dotted lines, respectively.}
        \label{fig:fig11}
\end{figure}

The DNS data from Lee and Moser (2015) \cite{lee15} and Pirozzoli (2024) \cite{piroz24} are used to generate figure \ref{fig:fig7}.  The trend reveals that, independent of $Re_\tau$, the viscous stress is less than $20\%$ of the total stress for $y^+ \gtrapprox 20$, which is just past the location of the normal stress peak found typically around $y^+ \approx 15$. In the following, the evaluation of the logarithmic and defect-power normal stress models is reconsidered in the region between the location of the normal stress peak and the overlap region based on both DNS and experimental data used by Nagib and Marusic (2024) \cite {nagmar24}.

Figure \ref{fig:fig7} shows the normal stress, $<u_1^+ u_1^+>$, versus $y^+$ for the various cases of $Re_\tau$ in the channel flow by Lee and Moser \cite{lee15} using different colors. With the same colors, the best fits of equation \ref{eq:001} are shown using the solid lines to represent the logarithmic trend. Similarly, the best fits of the defect-power trend are shown in dotted lines.  The fits are carried out in the region outside the location of the peak of the normal stress but much closer to the wall than the overlap region used by Nagib and Marusic (2024) \cite{nagmar24} and all other investigations of this subject.  To the right of the vertical black dashed line, the viscous stress is less than $20\%$ and in the middle of the fitting region, the viscous stress is approximately $5\%$.

\begin{figure}
      \centering
      \includegraphics[width=0.65\textwidth]{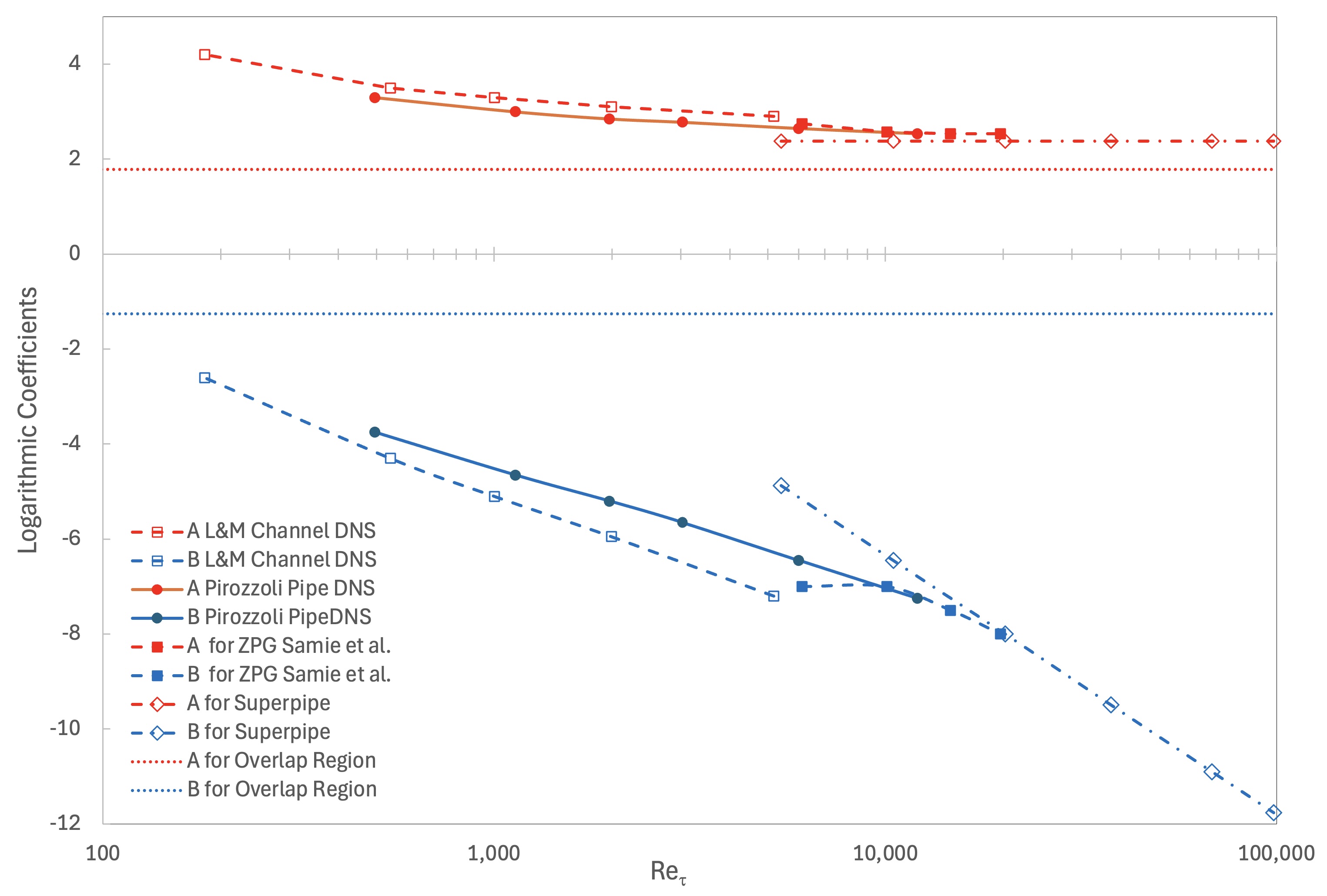}
        \caption{Variation with $Re_\tau$ of the parameters of the logarithmic model of equation \ref{eq:001} for DNS and experimental data in channel, pipe and ZPG boundary layer flows of figures \ref{fig:fig8}, \ref{fig:fig9}, \ref{fig:fig10} and \ref{fig:fig11}.}
        \label{fig:fig12}
\end{figure}

\begin{figure}
      \centering
      \includegraphics[width=0.65\textwidth]{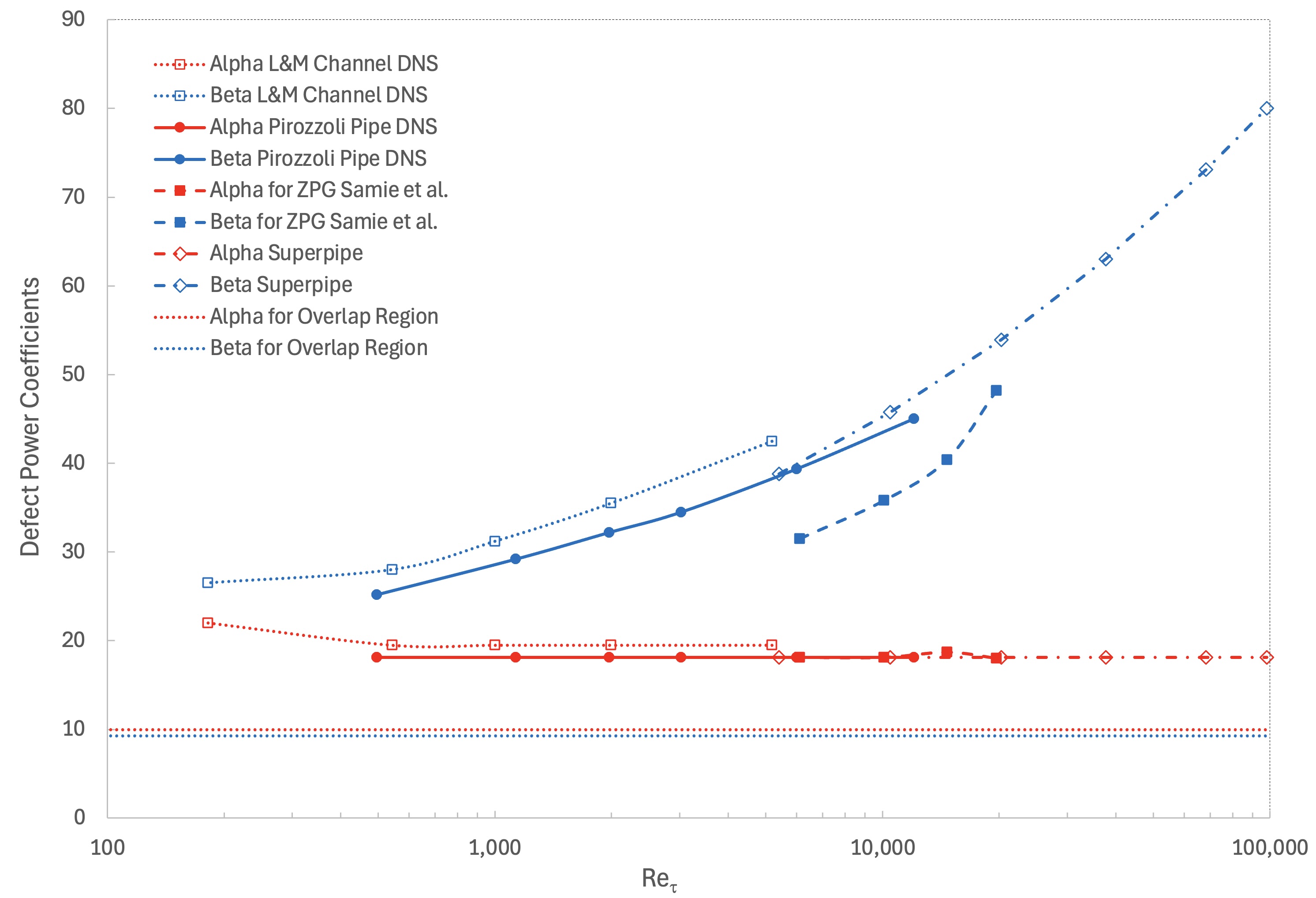}
        \caption{Variation with $Re_\tau$ of the parameters of the defect-power model of equation \ref{eq:002} for DNS and experimental data in channel, pipe and ZPG boundary layer flows of figures \ref{fig:fig8}, \ref{fig:fig9}, \ref{fig:fig10} and \ref{fig:fig11}..}
        \label{fig:fig13}
\end{figure}

Figure \ref{fig:fig8} presents a plot similar to figure \ref{fig:fig7} but for the DNS data of the pipe flow by Pirozzoli \cite{piroz24}. The values of the parameters $A_1$ and $B_1$ of equation \ref{eq:001} and $\alpha_1$ and $\beta_1$ of equation \ref{eq:002} for these fits are included in figures \ref{fig:fig12} and \ref{fig:fig13}. 

The experimental results in the ZPG boundary layers of Samie et al. (2018) \cite{sam18} and in the flow of pipes of Hultmark et al. (2012) \cite{hul12} are presented in the same way in figures \ref{fig:fig10}
 and \ref{fig:fig11}.  In these two figures, the best fit of the logarithmic tren and the defect-power trend in the overlap region reported by Nagib and Marusic (2024) \cite{nagmar24} are also included to contrast with the current results.

The values of the parameters $A_1$ and $B_1$ of equation \ref{eq:001} and $\alpha_1$ and $\beta_1$ of equation \ref{eq:002} for the best fits of both DNS and experimental data are summarized figures \ref{fig:fig12} and \ref{fig:fig13}, with the values of the same parameters when the logarithmic and defect-power trends are applied in the overlap region  between $y^+_{in} \gtrapprox 400$ and $Y_{out} \approx 0.45$.

\section{Conclusions}
Following recent evidence that even ZPG boundary layers do not exhibit a pure-log extended overlap region by Baxerras et al. (2024) \cite{bax24}, reconsideration of the Monkewitz and Nagib (2023)\cite{mon23} overlap equation \ref{eq:006} is advanced here. Examining a large volume of DNS data and some experiments in which accurate near-wall measurements are available, a revised representation of the extended overlap region is proposed by equation \ref{eq:007}. The significant differences between the two representations are incorporation of a pure-log behavior in part of the inner wall layer, and the separation between the inner layer and the overlap layer in the coefficient of the logarithmic term into $\kappa_{in}$ and $\kappa_o$, respectively.  In the physically small pure-log region under high $Re_\tau$ conditions, the logarithmic term coefficient in the mean velocity profile is universal for all wall-bounded flows and equal to $2.6$, or $\kappa_{in}$ in the range $0.38$ to $0.39$.  For the extended overlap region where the log+lin conditions are established, the coefficient of the log term $1/\kappa_o$ is not universal and depends on the flow geometry and its imposed pressure gradient, consistent with the conclusions of Baxerras et al. (2024) \cite{bax24} and Monkewitz and Nagib (2023)\cite{mon23}. 

In recent DNS data of pipe and channel flows, no evidence is found to establish a correspondence between a constant turbulence shear stress and the pure log regions of the mean velocity profile. It is very difficult to rely on experimental data for a similar confirmation.

Nagib and Marusic (2024) \cite{nagmar24} concluded that the defect-power model of equation \ref{eq:002} is in more agreement with experimental data from ZPG boundary layer and pipe flows than the logarithmic model of equation \ref{eq:001}. Similar agreement was demonstrated by Nagib et al. (2024) \cite{nag24} using DNS data for channel and pipe flows.  For these investigations and the entire literature on this popular topic, the assessment is made in the overlap region between inner and outer flows that have limited viscous effects and are essentially inviscid. This appears to be counterintuitive and deserves further attention. Here we evaluated the two models in a region closer to the wall but outside of the region dominated by viscous effects; that is, the ratio of viscous over total stresses greater than $20\%$. It is also perplexing that both an inviscid and a viscous model agree equally well with experiments and DNS data closer to the wall but just outside the viscous-dominated wall layer.

Both parameters established in this region for the two models using equation \ref{eq:001} ($A_1$ $\&$ $B_1$) and equation \ref{eq:002} ($\alpha_1$ $\&$ $\beta_1$) are different from those generally found in the overlap region farther from the wall. The dependence on Reynolds number of the four parameters is shown in figures \ref{fig:fig12} and \ref{fig:fig13} over the range $200 < Re_\tau < 100,000$, and the extracted values are consistent between the experimental measurements and the DNS results of flows from channel, pipe and ZPG boundary layer flows. 

Based on the wall-scaled eddies model leading to equation \ref{eq:001} and as confirmed by various experiments including Baars and Marusic (2020) \cite{baars2020part2} in a ZPG boundary layer, the value of the coefficient of the logarithmic term $A$ in the outer overlap region over a similar range of $Re_\tau$ is constant and represented by the dotted line in figure \ref{fig:fig12}. The higher values obtained for the coefficient $A$ closer to the wall may trend asymptotically to the same value or are more likely influenced by the contribution of viscous stress $5\%$ to $20\%$ in the range considered closer to the wall.  The apparent constant value extracted from the pipe data of Hultmark et al. (2012) \cite{hul12} is probably affected by the limited resolutions of the measuring sensors.

An important conclusion from examination of various models intended for the high limit of Reynolds numbers here and in the recent work of Nagib and Marusic (2024) \cite{nagmar24}, is that with all advances in facilities, measurement techniques, and computational modeling we remain far from reaching adequate $Re_\tau$ conditions.  We also must remain focused with great care on spatial resolution, accuracy, proper simulation conditions, and adequate convergence of computations; see, e.g., Hoyas et al. (2024) \cite{hoyas24}. For example, higher $Re_\tau$ experiments would expand the range of $y^+$ where high Reynolds number models for normal stress can be evaluated in region closer to wall.

\section{Acknowledgments}
The support of the Rettaliata Chair Professorship at Illinois Tech over many years is appreciated. Feedback on the manuscript by Ivan Marusic, Rahul Deshpande, Mitchell Lozier, and K. R. Sreenivasan was very valuable.



\section*{References}\label{ref}

\begin{thebibliography}{10}
\bibitem [{\citenamefont {Nagib}(2024)}]{nagaps24}%
  \BibitemOpen
  \bibfield  {author} {\bibinfo {author} {\bibfnamefont {H.}~\bibnamefont {Nagib}},\ }\bibfield  {title} {\bibinfo {title} {Invited talk: New fundamental developments in wall-bounded turbulence},\ }\href {https://doi.org/10.1175/1520-0450(1999)038<1576:ANSOFA>2.0.CO;2} {\bibfield  {journal} {\bibinfo  {journal} {Buletin of American Physical Society}\ }\textbf {\bibinfo {volume} {77th Annual Meeting of the Division of Fluid Dynamics}},\ \bibinfo {pages} {November 24–26, 2024; Salt Lake City, Utah} (\bibinfo {year} {2024})}\BibitemShut {NoStop}%
\bibitem [{\citenamefont {Monkewitz}\ and\ \citenamefont {Nagib}(2023)}]{mon23}%
  \BibitemOpen
  \bibfield  {author} {\bibinfo {author} {\bibfnamefont {P.}~\bibnamefont {Monkewitz}}\ and\ \bibinfo {author} {\bibfnamefont {H.}~\bibnamefont {Nagib}},\ }\bibfield  {title} {\bibinfo {title} {The hunt for the {K}ármán ‘constant’ revisited},\ }\href {https://doi.org/10.1017/jfm.2023.448} {\bibfield  {journal} {\bibinfo  {journal} {Journal of Fluid Mechanics}\ }\textbf {\bibinfo {volume} {967}},\ \bibinfo {pages} {A15} (\bibinfo {year} {2023})}\BibitemShut {NoStop}%
\bibitem [{\citenamefont {Baxerras}\ \emph {et~al.}(2024)\citenamefont {Baxerras}, \citenamefont {Vinuesa},\ and\ \citenamefont {Nagib}}]{bax24}%
  \BibitemOpen
  \bibfield  {author} {\bibinfo {author} {\bibfnamefont {V.}~\bibnamefont {Baxerras}}, \bibinfo {author} {\bibfnamefont {R.}~\bibnamefont {Vinuesa}},\ and\ \bibinfo {author} {\bibfnamefont {H.}~\bibnamefont {Nagib}},\ }\bibfield  {title} {\bibinfo {title} {Evidence of quasiequilibrium in pressure-gradient turbulent boundary layers},\ }\href@noop {} {\bibfield  {journal} {\bibinfo  {journal} {Journal of Fluid Mechanics}\ }\textbf {\bibinfo {volume} {987}},\ \bibinfo {pages} {R8} (\bibinfo {year} {2024})}\BibitemShut {NoStop}%
\bibitem [{\citenamefont {Nagib}\ and\ \citenamefont {Marusic}(2024)}]{nagmar24}%
  \BibitemOpen
  \bibfield  {author} {\bibinfo {author} {\bibfnamefont {H.}~\bibnamefont {Nagib}}\ and\ \bibinfo {author} {\bibfnamefont {I.}~\bibnamefont {Marusic}},\ }\bibfield  {title} {\bibinfo {title} {A method for evaluating relations of turbulent normal-stresses by experimental data over a wide range of reynolds numbers},\ }\href@noop {} {\bibfield  {journal} {\bibinfo  {journal} {arXiv preprint arXiv:2410.23669v2}\ } (\bibinfo {year} {2024})}\BibitemShut {NoStop}%
\bibitem [{\citenamefont {Nagib}\ and\ \citenamefont {Chauhan}(2008)}]{variations}%
  \BibitemOpen
  \bibfield  {author} {\bibinfo {author} {\bibfnamefont {H.~M.}\ \bibnamefont {Nagib}}\ and\ \bibinfo {author} {\bibfnamefont {K.~A.}\ \bibnamefont {Chauhan}},\ }\bibfield  {title} {\bibinfo {title} {Variations of von {K}\'arm\'an coefficient in canonical flows},\ }\href@noop {} {\bibfield  {journal} {\bibinfo  {journal} {Physics of Fluids}\ }\textbf {\bibinfo {volume} {20}},\ \bibinfo {pages} {101518} (\bibinfo {year} {2008})}\BibitemShut {NoStop}%
\bibitem [{\citenamefont {Nagib}\ \emph {et~al.}(2024)\citenamefont {Nagib}, \citenamefont {Vinuesa},\ and\ \citenamefont {Hoyas}}]{nag24}%
  \BibitemOpen
  \bibfield  {author} {\bibinfo {author} {\bibfnamefont {H.}~\bibnamefont {Nagib}}, \bibinfo {author} {\bibfnamefont {R.}~\bibnamefont {Vinuesa}},\ and\ \bibinfo {author} {\bibfnamefont {S.}~\bibnamefont {Hoyas}},\ }\bibfield  {title} {\bibinfo {title} {Utilizing indicator functions with computational data to confirm nature of overlap in normal turbulent stresses: logarithmic or quarter-power},\ }\href@noop {} {\bibfield  {journal} {\bibinfo  {journal} {Physics of Fluids 36, 075145}\ }\textbf {\bibinfo {volume} {36}} (\bibinfo {year} {2024})}\BibitemShut {NoStop}%
\bibitem [{\citenamefont {Monkewitz}(2023)}]{M23}%
  \BibitemOpen
  \bibfield  {author} {\bibinfo {author} {\bibfnamefont {P.}~\bibnamefont {Monkewitz}},\ }\bibfield  {title} {\bibinfo {title} {Reynolds number scaling and inner-outer overlap of stream-wise reynolds stress in wall turbulence},\ }\href@noop {} {\bibfield  {journal} {\bibinfo  {journal} {arXiv:2307.00612v3}\ } (\bibinfo {year} {2023})}\BibitemShut {NoStop}%
\bibitem [{\citenamefont {Marusic}\ and\ \citenamefont {Monty}(2019)}]{mar19}%
  \BibitemOpen
  \bibfield  {author} {\bibinfo {author} {\bibfnamefont {I.}~\bibnamefont {Marusic}}\ and\ \bibinfo {author} {\bibfnamefont {J.~P.}\ \bibnamefont {Monty}},\ }\bibfield  {title} {\bibinfo {title} {Attached eddy model of wall turbulence},\ }\href@noop {} {\bibfield  {journal} {\bibinfo  {journal} {Annual Review of Fluid Mechanics}\ }\textbf {\bibinfo {volume} {51}},\ \bibinfo {pages} {49} (\bibinfo {year} {2019})}\BibitemShut {NoStop}%
\bibitem [{\citenamefont {Samie}\ \emph {et~al.}(2018)\citenamefont {Samie}, \citenamefont {Marusic}, \citenamefont {Hutchins}, \citenamefont {Fu}, \citenamefont {Fan}, \citenamefont {Hultmark},\ and\ \citenamefont {Smits}}]{sam18}%
  \BibitemOpen
  \bibfield  {author} {\bibinfo {author} {\bibfnamefont {M.}~\bibnamefont {Samie}}, \bibinfo {author} {\bibfnamefont {I.}~\bibnamefont {Marusic}}, \bibinfo {author} {\bibfnamefont {N.}~\bibnamefont {Hutchins}}, \bibinfo {author} {\bibfnamefont {M.}~\bibnamefont {Fu}}, \bibinfo {author} {\bibfnamefont {Y.}~\bibnamefont {Fan}}, \bibinfo {author} {\bibfnamefont {M.}~\bibnamefont {Hultmark}},\ and\ \bibinfo {author} {\bibfnamefont {A.}~\bibnamefont {Smits}},\ }\bibfield  {title} {\bibinfo {title} {{Fully resolved measurements of turbulent boundary layer flows up to $Re_\tau=20000$}},\ }\href {https://doi.org/10.1017/jfm.2018.508} {\bibfield  {journal} {\bibinfo  {journal} {Journal of Fluid Mechanics}\ }\textbf {\bibinfo {volume} {851}},\ \bibinfo {pages} {391} (\bibinfo {year} {2018})}\BibitemShut {NoStop}%
\bibitem [{\citenamefont {Chen}\ and\ \citenamefont {Sreenivasan}(2022)}]{che22}%
  \BibitemOpen
  \bibfield  {author} {\bibinfo {author} {\bibfnamefont {X.}~\bibnamefont {Chen}}\ and\ \bibinfo {author} {\bibfnamefont {K.~R.}\ \bibnamefont {Sreenivasan}},\ }\bibfield  {title} {\bibinfo {title} {Law of bounded dissipation and its consequences in turbulent wall flows},\ }\href@noop {} {\bibfield  {journal} {\bibinfo  {journal} {Journal of Fluid Mechanics}\ }\textbf {\bibinfo {volume} {933}},\ \bibinfo {pages} {A20} (\bibinfo {year} {2022})}\BibitemShut {NoStop}%
\bibitem [{\citenamefont {Chen}\ and\ \citenamefont {Sreenivasan}(2023)}]{che23}%
  \BibitemOpen
  \bibfield  {author} {\bibinfo {author} {\bibfnamefont {X.}~\bibnamefont {Chen}}\ and\ \bibinfo {author} {\bibfnamefont {K.~R.}\ \bibnamefont {Sreenivasan}},\ }\bibfield  {title} {\bibinfo {title} {Reynolds number asymptotics of wall-turbulence fluctuations},\ }\href@noop {} {\bibfield  {journal} {\bibinfo  {journal} {Journal of Fluid Mechanics}\ }\textbf {\bibinfo {volume} {976}},\ \bibinfo {pages} {A21} (\bibinfo {year} {2023})}\BibitemShut {NoStop}%
\bibitem [{\citenamefont {Pirozzoli}(2024)}]{piroz24}%
  \BibitemOpen
  \bibfield  {author} {\bibinfo {author} {\bibfnamefont {S.}~\bibnamefont {Pirozzoli}},\ }\bibfield  {title} {\bibinfo {title} {On the streamwise velocity variance in the near-wall region of turbulent flows},\ }\href@noop {} {\bibfield  {journal} {\bibinfo  {journal} {Journal of Fluid Mechanics}\ }\textbf {\bibinfo {volume} {989}},\ \bibinfo {pages} {A5} (\bibinfo {year} {2024})}\BibitemShut {NoStop}%
\bibitem [{\citenamefont {Marusic}\ \emph {et~al.}(2013)\citenamefont {Marusic}, \citenamefont {Monty}, \citenamefont {Hultmark},\ and\ \citenamefont {Smits}}]{mar13}%
  \BibitemOpen
  \bibfield  {author} {\bibinfo {author} {\bibfnamefont {I.}~\bibnamefont {Marusic}}, \bibinfo {author} {\bibfnamefont {J.~P.}\ \bibnamefont {Monty}}, \bibinfo {author} {\bibfnamefont {M.}~\bibnamefont {Hultmark}},\ and\ \bibinfo {author} {\bibfnamefont {A.~J.}\ \bibnamefont {Smits}},\ }\bibfield  {title} {\bibinfo {title} {On the logarithmic region in wall turbulence},\ }\href@noop {} {\bibfield  {journal} {\bibinfo  {journal} {Journal of Fluid Mechanics}\ }\textbf {\bibinfo {volume} {716}},\ \bibinfo {pages} {R3} (\bibinfo {year} {2013})}\BibitemShut {NoStop}%
\bibitem [{\citenamefont {Hwang}\ \emph {et~al.}(2022)\citenamefont {Hwang}, \citenamefont {Hutchins},\ and\ \citenamefont {Marusic}}]{hwa22}%
  \BibitemOpen
  \bibfield  {author} {\bibinfo {author} {\bibfnamefont {Y.}~\bibnamefont {Hwang}}, \bibinfo {author} {\bibfnamefont {N.}~\bibnamefont {Hutchins}},\ and\ \bibinfo {author} {\bibfnamefont {I.}~\bibnamefont {Marusic}},\ }\bibfield  {title} {\bibinfo {title} {The logarithmic variance of streamwise velocity and k-1 conundrum in wall turbulence},\ }\href@noop {} {\bibfield  {journal} {\bibinfo  {journal} {Journal of Fluid Mechanics}\ }\textbf {\bibinfo {volume} {933}},\ \bibinfo {pages} {A8} (\bibinfo {year} {2022})}\BibitemShut {NoStop}%
\bibitem [{\citenamefont {Diwan}\ and\ \citenamefont {Morrison}(2021)}]{Diw21}%
  \BibitemOpen
  \bibfield  {author} {\bibinfo {author} {\bibfnamefont {S.~S.}\ \bibnamefont {Diwan}}\ and\ \bibinfo {author} {\bibfnamefont {J.~F.}\ \bibnamefont {Morrison}},\ }\bibfield  {title} {\bibinfo {title} {Intermediate scaling and logarithmic invariance in turbulent pipe flow},\ }\href@noop {} {\bibfield  {journal} {\bibinfo  {journal} {Journal of Fluid Mechanics}\ }\textbf {\bibinfo {volume} {913}},\ \bibinfo {pages} {R1} (\bibinfo {year} {2021})}\BibitemShut {NoStop}%
\bibitem [{\citenamefont {Hultmark}\ \emph {et~al.}(2012)\citenamefont {Hultmark}, \citenamefont {Vallikivi}, \citenamefont {Bailey},\ and\ \citenamefont {Smits}}]{hul12}%
  \BibitemOpen
  \bibfield  {author} {\bibinfo {author} {\bibfnamefont {M.}~\bibnamefont {Hultmark}}, \bibinfo {author} {\bibfnamefont {M.}~\bibnamefont {Vallikivi}}, \bibinfo {author} {\bibfnamefont {S.~C.~C.}\ \bibnamefont {Bailey}},\ and\ \bibinfo {author} {\bibfnamefont {A.~J.}\ \bibnamefont {Smits}},\ }\bibfield  {title} {\bibinfo {title} {Turbulent pipe flow at extreme reynolds numbers},\ }\href {https://doi.org/10.1103/PhysRevLett.108.094501} {\bibfield  {journal} {\bibinfo  {journal} {Physical Review Letters}\ }\textbf {\bibinfo {volume} {108}},\ \bibinfo {pages} {094501} (\bibinfo {year} {2012})}\BibitemShut {NoStop}%
\bibitem [{\citenamefont {Marusic}\ \emph {et~al.}(2015)\citenamefont {Marusic}, \citenamefont {Chauhan}, \citenamefont {Kulandaivelu},\ and\ \citenamefont {Hutchins}}]{mar15}%
  \BibitemOpen
  \bibfield  {author} {\bibinfo {author} {\bibfnamefont {I.}~\bibnamefont {Marusic}}, \bibinfo {author} {\bibfnamefont {K.}~\bibnamefont {Chauhan}}, \bibinfo {author} {\bibfnamefont {V.}~\bibnamefont {Kulandaivelu}},\ and\ \bibinfo {author} {\bibfnamefont {N.}~\bibnamefont {Hutchins}},\ }\bibfield  {title} {\bibinfo {title} {Evolution of zero-pressure-gradient boundary layers from different tripping conditions},\ }\href {https://doi.org/10.1017/jfm.2015.556} {\bibfield  {journal} {\bibinfo  {journal} {Journal of Fluid Mechanics}\ }\textbf {\bibinfo {volume} {783}},\ \bibinfo {pages} {379} (\bibinfo {year} {2015})}\BibitemShut {NoStop}%
\bibitem [{\citenamefont {Chen}\ and\ \citenamefont {Sreenivasan}(2021)}]{che21}%
  \BibitemOpen
  \bibfield  {author} {\bibinfo {author} {\bibfnamefont {X.}~\bibnamefont {Chen}}\ and\ \bibinfo {author} {\bibfnamefont {K.~R.}\ \bibnamefont {Sreenivasan}},\ }\bibfield  {title} {\bibinfo {title} {Reynolds number scaling of the peak turbulence intensity in wall flows},\ }\href@noop {} {\bibfield  {journal} {\bibinfo  {journal} {Journal of Fluid Mechanics}\ }\textbf {\bibinfo {volume} {908}},\ \bibinfo {pages} {R3} (\bibinfo {year} {2021})}\BibitemShut {NoStop}%
\bibitem [{\citenamefont {Lee}\ and\ \citenamefont {Moser}(2015)}]{lee15}%
  \BibitemOpen
  \bibfield  {author} {\bibinfo {author} {\bibfnamefont {M.}~\bibnamefont {Lee}}\ and\ \bibinfo {author} {\bibfnamefont {R.}~\bibnamefont {Moser}},\ }\bibfield  {title} {\bibinfo {title} {Direct numerical simulation of turbulent channel flow up to ${R}e_\tau\approx5200$},\ }\href@noop {} {\bibfield  {journal} {\bibinfo  {journal} {Journal of Fluid Mechanics}\ }\textbf {\bibinfo {volume} {774}},\ \bibinfo {pages} {395} (\bibinfo {year} {2015})}\BibitemShut {NoStop}%
\bibitem [{\citenamefont {Chauhan}\ \emph {et~al.}(2009)\citenamefont {Chauhan}, \citenamefont {Monkewitz},\ and\ \citenamefont {Nagib}}]{FDR09}%
  \BibitemOpen
  \bibfield  {author} {\bibinfo {author} {\bibfnamefont {K.}~\bibnamefont {Chauhan}}, \bibinfo {author} {\bibfnamefont {P.}~\bibnamefont {Monkewitz}},\ and\ \bibinfo {author} {\bibfnamefont {H.}~\bibnamefont {Nagib}},\ }\bibfield  {title} {\bibinfo {title} {Criteria for assessing experiments in zero pressure gradient boundary layers},\ }\href@noop {} {\bibfield  {journal} {\bibinfo  {journal} {Fluid Dynamics Research}\ }\textbf {\bibinfo {volume} {41}},\ \bibinfo {pages} {021404} (\bibinfo {year} {2009})}\BibitemShut {NoStop}%
\bibitem [{\citenamefont {Hoyas}\ \emph {et~al.}(2024)\citenamefont {Hoyas}, \citenamefont {Vinuesa}, \citenamefont {Schmidt},\ and\ \citenamefont {Nagib}}]{hoyas24}%
  \BibitemOpen
  \bibfield  {author} {\bibinfo {author} {\bibfnamefont {S.}~\bibnamefont {Hoyas}}, \bibinfo {author} {\bibfnamefont {R.}~\bibnamefont {Vinuesa}}, \bibinfo {author} {\bibfnamefont {P.}~\bibnamefont {Schmidt}},\ and\ \bibinfo {author} {\bibfnamefont {H.}~\bibnamefont {Nagib}},\ }\bibfield  {title} {\bibinfo {title} {Sensitivity study of resolution and convergence requirements for the extended overlap region in wall-bounded turbulence},\ }\href {https://doi.org/10.1103/PhysRevLett.108.094501} {\bibfield  {journal} {\bibinfo  {journal} {Physical Review Letters}\ }\textbf {\bibinfo {volume} {108}},\ \bibinfo {pages} {L082601} (\bibinfo {year} {2024})}\BibitemShut {NoStop}%
\bibitem [{\citenamefont {Baars}\ and\ \citenamefont {Marusic}(2020)}]{baars2020part2}%
  \BibitemOpen
  \bibfield  {author} {\bibinfo {author} {\bibfnamefont {W.~J.}\ \bibnamefont {Baars}}\ and\ \bibinfo {author} {\bibfnamefont {I.}~\bibnamefont {Marusic}},\ }\bibfield  {title} {\bibinfo {title} {Data-driven decomposition of the streamwise turbulence kinetic energy in boundary layers. {P}art2. {I}ntegrated energy and ${A_{1}}$},\ }\href@noop {} {\bibfield  {journal} {\bibinfo  {journal} {J. Fluid Mech.}\ }\textbf {\bibinfo {volume} {882}},\ \bibinfo {pages} {{A}26} (\bibinfo {year} {2020})}\BibitemShut {NoStop}%



\end{thebibliography}

\makeatletter
\providecommand \@ifxundefined [1]{%
 \@ifx{#1\undefined}
}%
\providecommand \@ifnum [1]{%
 \ifnum #1\expandafter \@firstoftwo
 \else \expandafter \@secondoftwo
 \fi
}%
\providecommand \@ifx [1]{%
 \ifx #1\expandafter \@firstoftwo
 \else \expandafter \@secondoftwo
 \fi
}%
\providecommand \natexlab [1]{#1}%
\providecommand \enquote  [1]{``#1''}%
\providecommand \bibnamefont  [1]{#1}%
\providecommand \bibfnamefont [1]{#1}%
\providecommand \citenamefont [1]{#1}%
\providecommand \href@noop [0]{\@secondoftwo}%
\providecommand \href [0]{\begingroup \@sanitize@url \@href}%
\providecommand \@href[1]{\@@startlink{#1}\@@href}%
\providecommand \@@href[1]{\endgroup#1\@@endlink}%
\providecommand \@sanitize@url [0]{\catcode `\\12\catcode `\$12\catcode
  `\&12\catcode `\#12\catcode `\^12\catcode `\_12\catcode `\%12\relax}%
\providecommand \@@startlink[1]{}%
\providecommand \@@endlink[0]{}%
\providecommand \url  [0]{\begingroup\@sanitize@url \@url }%
\providecommand \@url [1]{\endgroup\@href {#1}{\urlprefix }}%
\providecommand \urlprefix  [0]{URL }%
\providecommand \Eprint [0]{\href }%
\providecommand \doibase [0]{https://doi.org/}%
\providecommand \selectlanguage [0]{\@gobble}%
\providecommand \bibinfo  [0]{\@secondoftwo}%
\providecommand \bibfield  [0]{\@secondoftwo}%
\providecommand \translation [1]{[#1]}%
\providecommand \BibitemOpen [0]{}%
\providecommand \bibitemStop [0]{}%
\providecommand \bibitemNoStop [0]{.\EOS\space}%
\providecommand \EOS [0]{\spacefactor3000\relax}%
\providecommand \BibitemShut  [1]{\csname bibitem#1\endcsname}%
\let\auto@bib@innerbib\@empty

\bibliographystyle{unsrt}

\end{document}